# Extracting optical parameters of Cu-Mn-Fe spinel oxide nanoparticles for optimizing air-stable, high-efficiency solar selective coatings

Xiaoxin Wang*[†], Can Xu, [†] and Jifeng Liu**

Thayer School of Engineering, Dartmouth College, 15 Thayer Drive, Hanover, New Hampshire 03755, USA

†Co-first authors contributing equally to this work

Corresponding authors

*Xiaoxin.Wang@Dartmouth.Edu (X. Wang)

*Jifeng.Liu@Dartmouth.Edu (J. Liu)

**Abstract**

High-temperature Cu-Mn-Fe spinel-oxide nanoparticle solar selective absorber coatings are investigated experimentally and theoretically. A reliable, general approach to evaluate absorption coefficient spectra from the optical measurements of the nanoparticle-pigmented coatings is developed based on solving the inverse problem using four-flux-radiative method. The derived absorption properties of NP materials can be directly applied to predict the solar absorptance, optimize the nanoparticle-pigmented coatings, and analyze the thermal degradation, which agree well with the experimental results. The analysis reveals that the Cu-Mn-Fe spinel oxides are fundamentally indirect bandgap ranging from 1.7 to 2.1 eV, while iron-free $CuMn_2O_4$ is a direct bandgap material with $E_g$=1.84 eV. With the same coating thickness and nanoparticle load, the solar absorptance ranks in the order of $Mn_2O_3$ < $MnFe_2O_4$ < $CuFe_2O_4$ < $CuFeMnO_4$ < $CuMn_2O_4$. The optimized spray-coated iron-free $CuMn_2O_4$ NP-pigmented coating demonstrates a high solar absorptance of 97%, a low emittance of 55%, a high optical-to-thermal energy conversion efficiency of ~93.5 % under 1000x solar concentration at 750ºC, and long-term endurance upon thermal cycling between 750°C and room temperature in air. The optical parameter analysis approach can be easily extended to other material systems to facilitate the searching and optimizing high-temperature pigmented-solar selective coatings.

**Highlight:**

**1.** Develop a general approach to extract the absorption coefficients and refractive indices of pigment nanoparticles from the optical measurements based on solving the inverse problem of the four-flux-radiative method.

**2.** Investigate and search for highly absorbing, high temperature Cu-Mn-Fe oxide pigmented solar selective coatings in a systematic way. With the same coating thickness and nanoparticle load, the pigment material performs in the order $Mn_2O_3 < MnFe_2O_4 < CuFe_2O_4 < CuFeMnO_4 < CuMn_2O_4$ in terms of solar absorptance

**3** Demonstrate iron-free $CuMn_2O_4$ coating with a high solar absorptance of 96.9%, low emittance of 55 %, and a high thermal efficiency of 93.5% under 1000x solar concentration at 750ºC that also endures long-term thermal cycling between 750 ºC and room temperature in air.

1. **Introduction**

Concentrating solar power (CSP) systems utilize optical components to collect and convert solar energy to thermal energy and then power heat engines to generate electricity. The widely used thermal energy storage in CSP systems allows the solar energy to be dispatched on demand,[1] providing a great advantage over other non-dispatchable renewable energy sources such as wind power and solar photovoltaic (PV) power. In order to reduce the levelized cost of energy (LCOE) of Generation 3 CSP systems towards 50% power efficiency, solar selective absorber coatings are required to possess long-term thermal stability at high temperatures ≥ 750 ºC in air.[2] Cost analysis reveals that durable oxidation-resistant solar selective coatings with solar absorptance $\alpha_{solar} \geq 95\%$ and thermal emittance $\varepsilon \leq 60\%$ can guarantee a reduction of the LCOE of CSP plant up to 12%.[3] Under this guideline, low-cost nanoparticle (NP)-pigmented solar absorbers with high solar absorptance and moderate spectral selectivity is advantageous over highly selective yet more costly and less thermally stable multilayer cermet coatings.

A very attractive nanoparticle (NP) pigment candidate for solar selective coatings is the mixed transition metal spinel oxides with a general formula $AB_2O_4$ (A, B = metal) because of their diverse properties and wide availability in versatile applications as electrodes[4], catalysts[5], magnetic materials[6,7]. Their tunable optical properties and high-temperature stability in air are especially suitable for solar absorber pigments [8-18]. These include Co oxide[8,9], Cu-Co oxide[10,11], Mn-Co oxide[10], Cu-Mn oxide[12], Cu-Ni-Co oxide[13], Cu-Co-Mn oxide[14], Cu-Cr-Mn oxide[15], and Cu-Mn-Fe oxide[16,17,18]. In our previous work, we have demonstrated air-stable $MnFe_2O_4$ spinel oxide NP-pigmented solar selective coatings with a high solar absorptance α ~93% and a thermal emittance ε ~55% for ~90% optical-to-thermal conversion efficiency ($\eta_{therm}$) using a small load of solar-absorbing transition metal oxide nanoparticles (10.5 vol. %).[19] Based on our established quantitative approach and experimental findings, the solar absorptance of pigmented coatings is realistically determined by the product of coating thickness ($d$) and NP volume concentration ($f$), which is flexible for design and optimization of solar absorbing coatings.[20] Furthermore, the appropriate selection of NP oxides pigments can conveniently tune and maximize the solar absorptance and solar selectivity, which is inherent to the NP materials, in contrast to strict control of layer thickness (~nm) to take advantage of the interference effect in the multilayer cermet coatings.[21]

However, there are still a couple of challenges in further optimizing these spinel-oxide NP solar coatings to approach the theoretical efficiency limit of 98%.[20] (1) The fundamental optical parameters of spinel oxide NPs are largely unavailable. The available data are only limited to some particular compositions in the form of bulk single crystals or thin films.[22, 23] Furthermore, defects, cation inversion and substitution on A and B sites of the spinel lattice, and non-stoichiometry due to synthesis methods may lead to a large deviation in the optical property of spinel NPs pigments (10-50 nm in diameter) from that of the bulk material [22,24,25,26]. Thus, it is important to develop a reliable method to derive wavelength-dependent effective absorption coefficient of the NPs from the measured optical spectra of the NP-pigmented composites. (2) The origin of the prominent anomalous sub-band-gap absorption in the near infrared (NIR) regime of the solar spectrum remains unclear in literature, which is important in order to optimize the solar selectivity. Mechanisms range from charge transfer between bivalence-metal ions occupying distorted octahedral or tetrahedral sites (e.g. in $Mn_xFe_{3-x}O_{4+\gamma}$ single crystals[22]) to Urbach tail absorption (e.g. in $CuCoO_x$)[27] to chemical substitution and partially inverted configuration between tetrahedral A sites and octahedral B sites.[28] Therefore, a more accurate understanding of the electronic transitions in spinel nanoparticles should be developed by studying and comparing more compounds of the same family.

Recently, we have investigated $CuMn_{2-x}Cr_xO_4$ spinel oxide NP-pigmented solar selective coating system and demonstrated high thermal efficiency >94% in air at 750 ºC.[29] This initial success opens a large exploration space to further understand the impact of transition metal cationic species and their lattice sites on the performance of high-temperature solar selective absorbers. In this paper, a group of spinel Cu-Mn-Fe oxide NPs are systematically investigated as efficient high-temperature solar selective absorbers. This is also scientifically motivated by a comparison of Fe vs. Cr alloying into $CuMn_2O_4$ NPs to gain more understanding, considering that Fe and Cr are on either side of Mn in the periodic table. A reliable, general approach to extract the absorption coefficient spectra from the optical measurements of the NP-pigmented coatings is developed based on solving the inverse problem using four-flux-radiative method. These derived effective scattering and absorption cross sections of synthesized NPs are then directly input into the four-flux radiative model to optimize the pigmented solar selective coatings. The optimized $CuMn_2O_4$ pigmented solar selective coating on the Inconel substrate demonstrates a high solar absorptance α =96.9% and a low thermal emittance ε =55.0% at 750 °C under 1000x solar concentration. This performance represents one of the highest efficiencies (93.5%) under this service condition and exceeds the requirements for Generation 3 CSP systems. The analysis reveals that iron-free $CuMn_2O_4$ spinel is a direct bandgap material with $E_g$=1.84 eV, while the spinel group of Cu-Mn-Fe are fundamentally indirect bandgap between 1.7~2.1 eV. Cyclic thermal stability testing between 750ºC and room temperature shows that the crystalline phase and cationic distribution of Cu-Mn-Fe spinel oxides are generally very durable, though some low valence states of transition metals might be oxidized to high valence states gradually and degrade the solar absorptance. Iron-free CuMn$_2$O$_4$ (synthesized Syn42 NP) exhibits the best long-term thermal sustainability for long-term thermal cycling, each cycle comprising 12h at 750°C and 12 h cooling to room temperature in air. The optical spectra curve-fitting and modeling approach developed here to extract the fundamental optical parameters of spinel NPs is applicable to other NP-pigmented solar coating systems to further optimize the efficiency.

## 2. Experimental and Modeling Methods

### 2.1 NP synthesis and spray coating deposition

Cu-Mn-Fe oxide NPs were synthesized by a co-precipitation method. Designed amounts of Cu$^{2+}$, Mn$^{2+}$ and Fe$^{3+}$ ions from Cu(NO$_3$)$_2$, Mn(NO$_3$)$_2$ and FeCl$_3$ were mixed with 100 mL deionized water to form solutions. After mechanically stirring, NaOH solution was added dropwise to the homogeneous mixture to adjust the pH value to 12 to facilitate the co-precipitation reaction. The precipitated NP were then collected after centrifugation, and washed repeatedly by deionized water to remove the excessive ions. Further drying, recrystallization and grinding steps were followed. As a comparison, commercially available MnFe$_2$O$_4$ NPs (purity: 99.99%, average diameter: 28 nm) and Mn$_2$O$_3$ NPs (purity: 99.99%, average diameter: 30 nm) were purchased from U.S. research Nanomaterials Inc.

To fabricate NPs-pigmented silicone solar selective coatings, synthesized NPs were well dispersed in xylene-diluted silicone matrix Bluesil through ultrasonic bath to form uniform precursors with different viscosities. The spray coatings were performed using a Houseales Mini sprayer (10 ml) on quartz substrates and high temperature Ni-based super-alloys Inconel 625 sheet coupons (1 in. × 1 in. square) heated to 80-120 °C by a hot plate. Here Ni-based super-alloy Inconel is chosen due to its good high-temperature mechanical behavior, oxidation resistance, and corrosion resistance. Multiple sprays were carried out to achieve the optimal coating thickness at a given NP pigment concentration. These samples were annealed at 750 ºC for 24 hours and then cooled down overnight for the following characterizations. All thermal cycling tests were performed in a box furnace in air. Each cycling test includes 12 h at 750 °C and 12 h at room temperature to mimic the day-night cycles in practical applications.

## 2.2 Characterization techniques

X-ray diffraction (XRD, Rigaku 007 X-ray Diffractometer, Cu K$\alpha$1 line, $\lambda$=0.15406 nm operating at 40 kV/300 mA and a scanning rate of 2°/min from 10° to 80°) was conducted to reveal information about crystal structure, phase weight percentage, average NP size, and cationic distribution of as-synthesized or annealed NPs in the form of powder or embedded in the coatings. High-resolution transmission electron microscopy images (TEM) were utilized to characterize the morphology, size and crystal structure of as-synthesized nanoparticles. Scanning electron microscopy (SEM, FEI XL-30 ESEM FEG, 20 kV, secondary electron (SE) mode) was performed to study the surface morphology and coating thickness. Energy dispersive spectroscopy (EDS, EDAX Si (Li) detector with Genesis software) was carried out to detect the chemical composition and elemental distribution in pigmented coatings. Raman spectra were recorded at room temperature using a confocal Raman imaging system and a laser radiation source operating at a wavelength of 532 nm.

The reflectance spectra in the wavelength range of $\lambda = 0.3 \sim 2.5$ µm were obtained by using a Jasco V-570 ultraviolet/visible/near infrared (UV/Vis/NIR) spectrometer equipped with a Jasco ISN-470 integrating sphere to capture both specular and diffuse reflection. The reflectance spectra in the mid infrared (MIR) region ($\lambda$=2.5 ~20 µm) for thermal emittance measurement was recorded with a Jasco 4100 Fourier transformation IR (FTIR) spectrometer equipped with a Pike IR integrating sphere in the range from 400 to 4000 cm$^{-1}$.

## 2.3 Four-flux radiative modeling and solution of the inverse problem

Four-flux radiative method is used to model the optical response of the NP-pigmented coatings and derive the optical properties of nanoparticle material from optical measurements with a Matlab curve fitting method. The details of four-flux radiative theory can be found in Refs. 19-20. Firstly, the cross sections of scattering ($C_{sca}$) and absorption ($C_{abs}$) of spherical NPs with a given average size were calculated using Lorentz-Mie theory (Mieplot[30]) based on the relative refractive index of NPs ($n_0+ik_0$) to that of the matrix material of the coating ($n_m+ik_m$). The absorption (K) and scattering coefficients (S) of the pigmented coating are calculated by K=$C_{abs}$*f/V and S=$C_{sca}$*f/V by taking into account the NP volume $V$ and NP volume fraction $f$. If coatings consist of NPs with various sizes and crystalline phases, the effective K and S of the coatings are defined as $K = \sum_{i=1}^{n} C_{abs,i} * f_i * V_i$ and $S = \sum_{i=1}^{n} C_{sca,i} * f_i * V_i$, respectively. Then the effective K, S of the coatings derived from the experiments or K, S values calculated from $C_{abs}$ and $C_{sca}$, along with the coating thickness, are input to the four-flux radiative models to calculate the reflectance R and transmittance T. For the coatings on the metal substrate such as Inconel, the light transited through the coating should be absorbed completely by the metal (T=0), thus only reflectance R is required.

In order to extract the complex refractive index of the NP material (n+ik) from the measured R and T by solving the inverse problem, two iterative loops of data fitting are introduced as shown in Fig. 1. One is based on the four-flux radiative method to derive effective K, S values of the coatings from the measured R, T, while the other loop uses the analytical Mie scattering to obtain the new refractive index of the NP material from the K, S. These loops are iterated until self-consistency is reached between the experimental and theoretical optical spectra. The derived optical parameters (i.e wavelength-dependent n and k) are then input to the four-flux radiative model to design the spinel oxide NP pigmented solar selective coatings, taking into account multiple scattering. The optical-to-thermal conversion efficiency |therm of the solar receiver is used as figure of merit (FOM) to optimize solar selective coating design, given by Equation 1

$$\eta_{therm} = FOM = \frac{\int_0^\infty (1-R(\lambda))I(\lambda)d\lambda - \frac{1}{C}\left[\int_0^\infty (1-R(\lambda))B(\lambda,T)d\lambda\right]}{\int_0^\infty I(\lambda)d\lambda} = \alpha_{solar} - \frac{\varepsilon\sigma T^4}{CI_{solar}} \quad (1)$$

Here $I(\lambda)$ is the AM 1.5 solar spectral radiance at wavelength $\lambda$; $I_{solar} = 1000\ W/m^2$ is the solar power density integrated from the spectral radiance $I(\lambda)$, $B(\lambda,T)$ is the spectral blackbody thermal emission at wavelength $\lambda$ and temperature $T$, $R(\lambda)$ is spectral reflectance, calculated with four-flux model. $\alpha_{solar}$ is the overall spectrally normalized solar absorptance, $\varepsilon$ is the overall thermal emittance at temperature $T$, $\sigma = 5.67 \times 10^{-8}\ \frac{W}{m^2 K^4}$ is the Stefan-Boltzmann constant, and $C$ is solar concentration ratio. In this paper, $T$=750°C=1023 K, and $C$=1000. The spectral range for the integrals is 300 nm-16 μm.

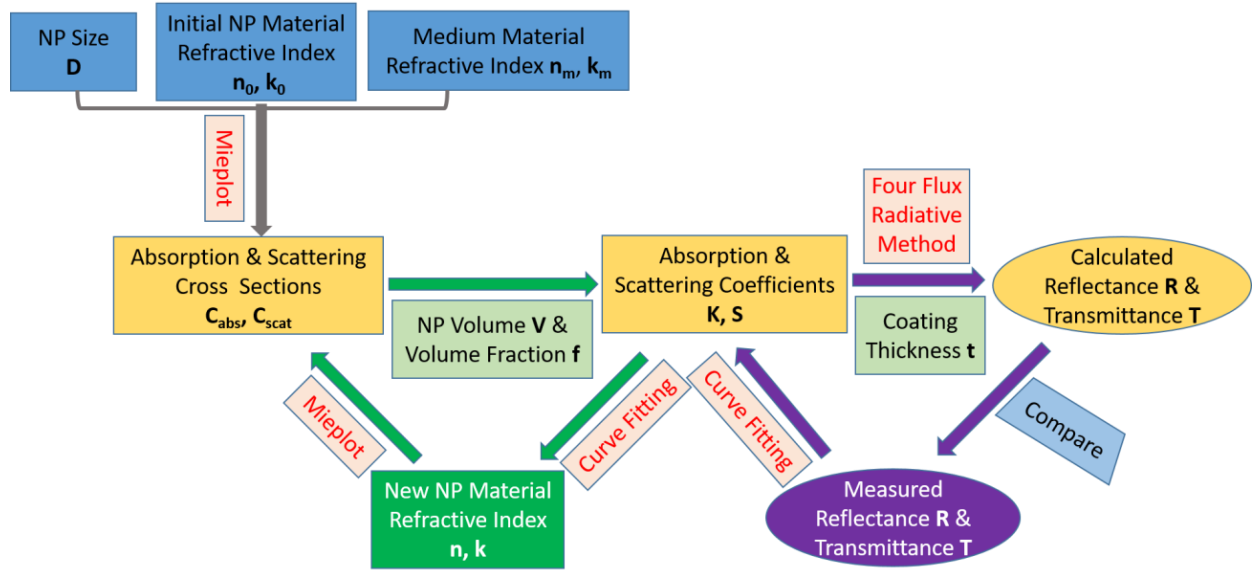

**Fig. 1** Flow of extracting the complex refractive index of NPs based on iterative, self-consistent Mie scattering theory and four-flux radiative method.

3. **Results and Discussion**

   **3.1 Characterization of NP and NP-pigmented coatings**

A group of Cu-Mn-Fe oxide NPs are synthesized by co-precipitation method and characterized by XRD, TEM and Raman spectroscopy. The crystalline phases and compositions are dependent on the starting material ratios of Cu:Mn:Fe, as summarized in Table 1 and detailed in the Supporting Information. As an example, the TEM and XRD data of Syn24 are shown in Fig. 2. The data for other samples are summarized in Figs. S1-S3 of the Supporting Information. Statistical analyses on TEM images of as-synthesized NPs (Fig. 2a and Fig. S1 (a)-(d)) show it is reasonable to use an average NP diameter D~50 nm for the optical spectra curve-fitting and theoretical modeling in the later context, as also indicated in Table 1. Both selected area electron diffraction patterns (Fig. S1(e)(f)) and XRD spectra (Fig. 2 and Fig. S2) reveal the co-existence of $Mn_2O_3$ and spinel oxide phases in the as-synthesized NPs due to phase separation for nonstoichiometric samples. In fact, a secondary $Mn_2O_3$ phase is inevitably formed along with cubic spinels when the ratio Cu:Mn ≤1.2 in $Cu_xMn_{3-x}O_4$ according to phase diagram,[31] which also holds true for Cu-Mn-Fe oxide family here. Furthermore, when XRD (311) peaks of spinel Cu-Mn-Fe oxide NPs are examined closely (Fig. 2c and Fig. S2(c)), each consists of two peaks for most annealed samples, which are ascribed to $CuFe_2O_4$ (~35.5°) and $CuMn_2O_4$ (~35.8°), respectively. Generally speaking, $CuMn_2O_4$ dominates the spinel phase for a starting material ratio Mn:Fe > 4:1, such as Syn31, 33, and 42. Otherwise, a higher or even dominant percentage of $CuFe_2O_4$ spinel phase exists, such as Syn 23-25. Syn29 and 35 have mixed $CuMn_2O_4$ and $CuFe_2O_4$. One particular case is Syn26 (Cu:Mn:Fe=1:6:2) with a single dominant (311) peak at 35.6°, matching well with the standard pattern of cubic $CuFeMnO_4$ (ICDD-PDF No.20-03588) spinel structure[32].

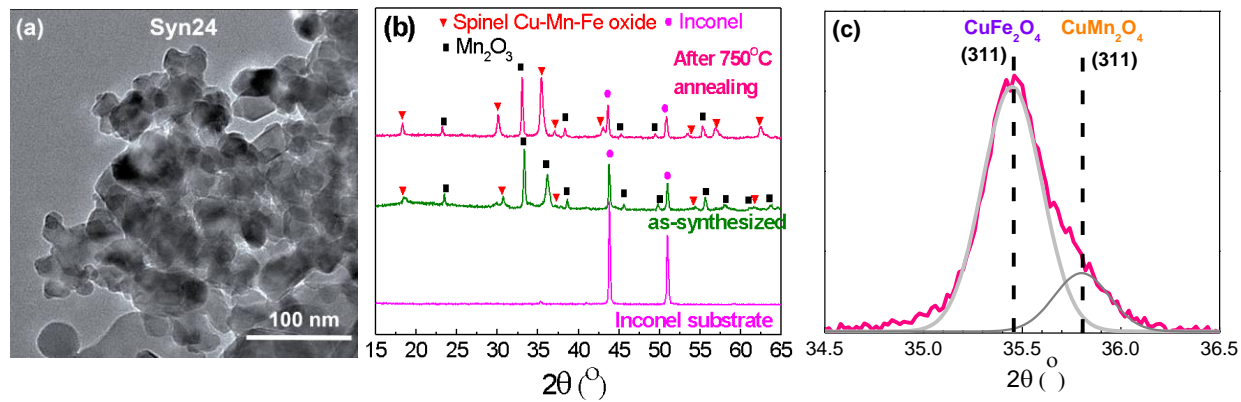

**Fig. 2** Cu-Mn-Fe oxide NPs Syn24: (a) A TEM image of as-synthesized NPs showing a region dominated by relatively small $CuFe_2O_4$ NPs. Larger $Mn_2O_3$ NPs from the same synthesis are shown in Fig. S1(b). (b) XRD spectra of as-synthesized NPs compared to those annealed at 750°C for 24 hours in air. The XRD data of the Inconel substrate is also shown as a reference. (c) Further deconvolution of (311) peaks of the annealed samples shown in (b), indicating co-existence of $CuFe_2O_4$ and $CuMn_2O_4$ spinel oxides.

**Table 1.** A list of Cu-Mn-Fe oxide NP samples investigated in this study. The last 3 columns apply to the NPs after annealing for 24h at 750ºC

| NP No. | Cu:Mn:Fe | TEM NP Size (nm)** | XRD NP-24h Size (nm) | Spinel Lattice Constant (Å) | Spinel Weight Percentage |
|---|---|---|---|---|---|
| Syn23 | 1:0.5:1.5 1:0.6:1.7* | 41.5±11.9 | 30 (sf); | 8.387 | 100% |
| Syn24 | 1:3:1 1:2.7:1.2* | 49.5±11.3 | 32 (sf); 47 (m) | 8.382*** | 46.3% |
| Syn25 | 1:1:3 1:1.2:3.2* | 49.2±8.5 | 24 (sf); 50 (m) | 8.382*** | 69.0% |
| Syn26 | 1:6:2 1:6.9:2.7* | 53.8±14.1 | 28(sfm); 40(m) | 8.376 | 71.2% |
| Syn29 | 1:4:1 1:3.9:0.9* | 48.0±9.1 | 52 (sf); 31 (sm); 54 (m) | 8.370/8.291 | 30.2% |
| Syn31 | 1:6:1 | - | 49 (sm); 58 (m) | 8.289 | 43.3% |
| Syn33 | 1:3:0.5 | - | 21 (sm); 50 (m) | 8.300 | 29.8% |
| Syn35 | 1:3:2 | - | 34 (sf); 42 (sm); 56 (m) | 8.374/8.295 | 82.3% |
| Syn42 | 1:2:0 | - | 31 (sm); 49 (m) | 8.305 | 70.9% |
| $Mn_2O_3$ | Purchased | 30 | - | - | 0% |
| $MnFe_2O_4$ | Purchased | 28 | - | - | 99.99% |

*EDS measured element ratio of Cu:Fe:Mn;
**sf**: spinel phase dominated by $CuFe_2O_4$; **sm**: spinel phase dominated by $CuMn_2O_4$;
**sfm**: spinel $CuFeMnO_4$; **m**: $Mn_2O_3$.
** As-synthesized  *** Lattice constant of the dominant sf phase.

Raman spectroscopy is used to further confirm the phases identified from XRD analysis, as shown in Fig. S3. Due to the strong absorption in the visible range, Raman signals from Syn24 and Syn42 are relatively weak. However, vibrational frequencies of the modes can be identified by multi-peak fitting. The Raman data provides the evidence that annealed Syn24 has mixed phases of $CuFe_2O_4$ and $Mn_2O_3$, and Syn42 has a dominant $CuMn_2O_4$ phase, in agreement with the XRD analysis.

Coatings comprising as-synthesized Cu-Mn-Fe oxide NPs dispersed in silicone matrix are deposited on quartz or Inconel substrates by a spray-coating method. Coatings are then treated at 750 °C for 24 hours to stabilize the cubic spinel phase of NPs and meanwhile improve the coating adhesion to the substrate. SEM images of Syn24-pigmented coating on Inconel are shown in Fig.S4. No cracks or warps are found on the coating surfaces. Coating thickness is ~8.5 µm estimated from the cross section image in Fig.S4 (c). The volume concentration of NP is ~13% calculated from the starting weight ratio of NPs and solid content of the silicone resin.

### 3.2 Extraction of Optical Parameters of NP-pigmented Coatings on Quartz

Following the flow in Fig. 1, the optical properties of synthesized Cu-Mn-Fe oxides NPs and commercial NPs ($Mn_2O_3$ and $MnFe_2O_4$) are determined inversely from the measured reflectance R and transmittance T of the corresponding pigmented coatings on quartz substrate. Fig. 3 shows the optical data of Syn29, Syn24, and Syn42, representing a trend of increasing weight percentages of spinel vs. $Mn_2O_3$ NPs (30%, 50% and 71%, respectively, as shown in Table 1). In Fig. 3(a) the measured absorptance spectra (A=1-R-T) of synthesized and commercial NP-pigmented coatings are nicely reproduced by the four flux radiative modeling using their derived optical parameters. Clearly, a full consistency is reached through the iteration loops in Fig. 1 for extracting the wavelength dependent complex refractive indices. Therefore, the approach developed here for solving the inverse problem of the four-flux method is very reliable. The correspondingly determined effective absorption coefficient of pigmented-coatings and derived refractive indices (both n and k) of NP materials are shown in Fig. 3(b) and Fig. 3(c), respectively, while Fig. 3(d) shows the cross sections of scattering ($C_{sca}$) and absorption ($C_{abs}$) for ~50 nm-diameter NPs based on Mie scattering theory using refractive indices in Fig.3(c).

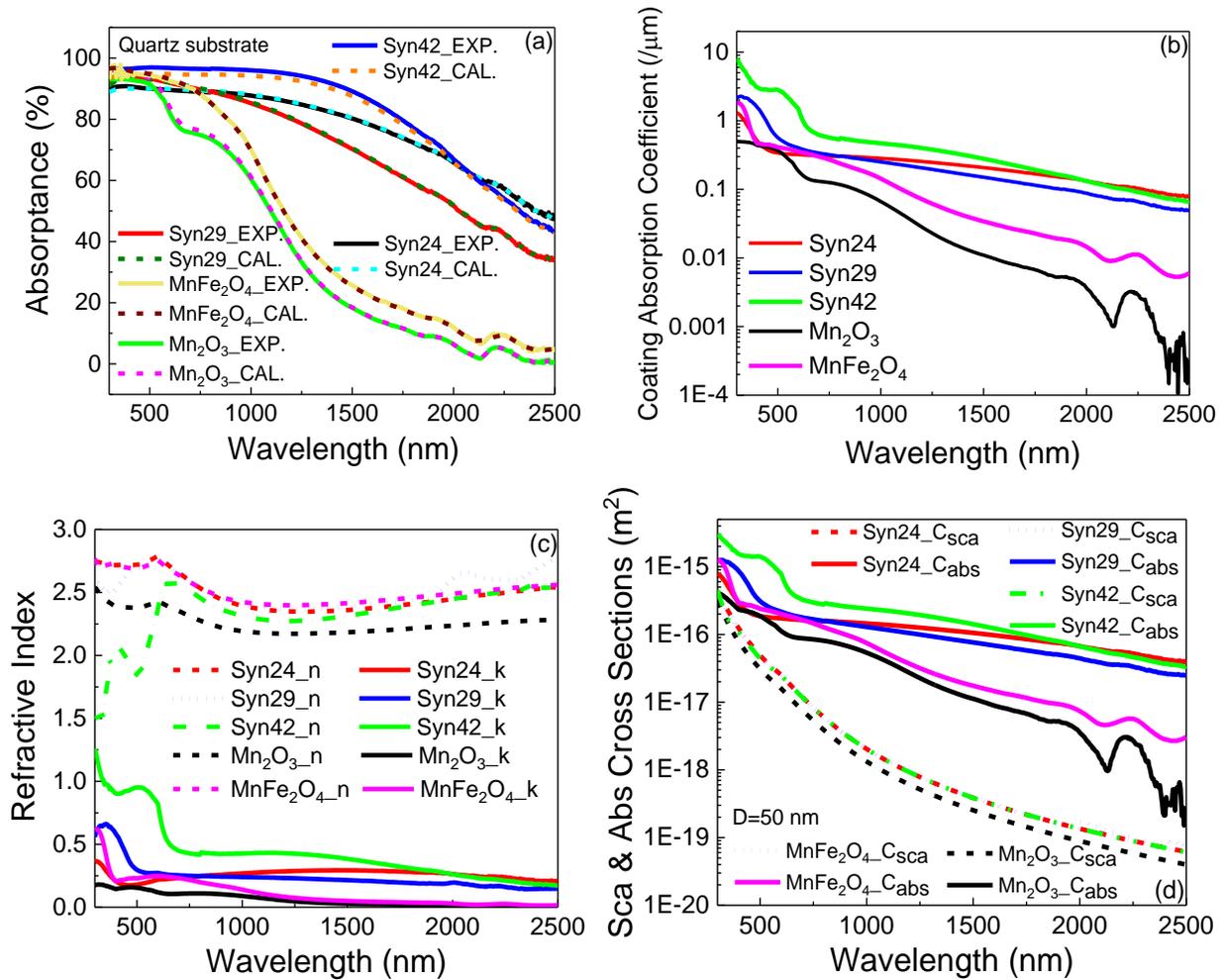

**Fig. 3** (a) Experimentally measured and theoretically modeled spectral absorptance of home-synthesized Syn24, Syn29 and Syn42 and commercial $Mn_2O_3$, $MnFe_2O_4$ NP-pigmented coatings on quartz substrate, showing excellent self-consistency is achieved after the iteration loops shown in Fig. 1. The extracted optical parameters from the iterative fitting are shown in (b)-(d). (b) The correspondingly derived effective absorption coefficients of the NP-coatings with different NP materials. (c) The corresponding derived effective refractive indices of NP materials, including both the real part n and the imaginary part k (extinction coefficient). (d) Calculated scattering and absorption cross sections of NPs with a size of 50 nm based on the refractive index data in (c).

In terms of solar absorption, the absorption curves of commercial $Mn_2O_3$ and $MnFe_2O_4$ NP-coatings in Fig. 3(a) roll off fast from wavelength > 600 nm in a similar way, though $MnFe_2O_4$-coating absorbs 5% more solar light than $Mn_2O_3$-coating. As spinel phases are incorporated, the absorption spectrum is clearly extended to longer wavelengths to better cover the solar spectrum, as shown in the curves for Syn 29, 24, and 42 (with spinel phase weight percentage increasing from 30% to 50% to 70%). Though Syn24-coating demonstrates a slightly smaller absorption at the wavelengths $\lambda$ < 800 nm than Syn29, it absorbs notably more infrared light than Syn29-coating in the wavelength range of 800~2500 nm. Overall, the iron-free Syn42 sample exhibits the highest absorption at $\lambda$<2000 nm despite of a slightly small absorption than Syn24 at $\lambda$> 2000 nm.

As far as the influence of $Mn_2O_3$ phase is concerned, we found that the effective absorption coefficients of NP-pigmented coatings incorporating spinel phases are 5-50x that of the reference $Mn_2O_3$-pigmented coating in Fig. 3 (b). Therefore, $Mn_2O_3$ phase make insignificant contribution to the solar absorption of synthesized NP-coatings, particularly in the wavelength range of 800~2500 nm. In other words, the curves in Fig. 3 (b) actually reflect the absorption trends and features of Cu-Mn-Fe spinel oxides in the synthesized NPs. The derived absorption coefficients of spinel NP materials excluding the effect of $Mn_2O_3$ will be presented later in Fig. 6.

Fig. 3(c) further shows that the real part of refractive indices across these samples are almost constant at n~2.3 at $\lambda$= 800-2500 nm. The variation in the real refractive indexes in the UV/Vis region at shorter wavelengths is associated with the strong band-to-band transitions seen in the absorption spectra (Fig.3(a)). As compared with-$Mn_2O_3$ NP pigmented coating, the spinel NP-containing coatings exhibit notable features in their imaginary parts of refractive indices: an absorption edge/peak in the UV/vis regime followed by a broad absorption band in the infrared band. The absorption edges/peaks are shifted with the spinel composition. Clearly, Syn24 (CuFe$_2$O$_4$ dominated), Syn29 (CuFe$_2$O$_4$+CuMn$_2$O$_4$) and Syn42 (CuMn$_2$O$_4$ dominated) are arranged in an ascending order of redshift in UV/Vis absorption spectra due to the corresponding bandgaps modifications summarized in Table 2, which will be further detailed in the analyses of Fig. 4 and Fig. 6 later.

Last but not least, Fig. 3(d) shows that the absorption cross sections of the spinel NPs is 10~100 times higher than the scattering cross sections for a NP diameter of 50 nm. Additionally, the absorption cross sections vary between different synthesized NPs, in contrast to the scattering cross sections insensitive to the NP materials discussed here. This is because the scattering cross-section is mainly determined by the real part of the refractive index, n, which are similar across different NPs as shown in Fig. 3(c). Furthermore, these effective absorption and scattering cross sections of mixed phases in synthesized NPs will be used to optimize these NP-pigmented solar absorbing coatings in Section 3.5.

### 3.3 Band-to-band and d-shell optical transitions of spinel NPs

**3.3.1 Effective absorption coefficient spectra of NPs comprising both spinel and Mn$_2$O$_3$ phases**: The absorption coefficients of NP materials, independent on NP size, can be calculated as α=4*pi*k/λ, where λ is the wavelength, and k is the imaginary part of the refractive index. Fig. 4 reveals the details of bandgap energy determination from the absorption coefficients using Tauc plots. The direct bandgap and indirect bandgap can be obtained by linear fit of (αE)$^2$ and (αE)$^{1/2}$ vs. photon energy E=hν, respectively. α-Mn$_2$O$_3$ has three direct gaps at 1.20 eV, 1.97 eV and 2.96 eV (Fig. 4(a)), in agreement with the reported 1.20 eV fundamental gap[33]. Fig. 4(b) and (c) respectively show the effective direct and indirect band gaps of the NPs (comprising spinel oxides and Mn$_2$O$_3$) in Syn24, Syn29 and Syn42. As mentioned earlier and further shown in Fig. 4a, the absorption is mainly contributed by spinel NPs rather than Mn$_2$O$_3$ NPs since the latter has much lower absorption coefficient. Syn42 with dominant cubic spinel $CuMn_2O_4$ phase is a direct band gap material with $E_g$=1.87 eV. Syn29, Syn24 and commercial $MnFe_2O_4$ NPs are fundamentally indirect, with bandgaps of 1.74 eV, 1.93 eV and 2.62 eV, respectively. All NPs shown in Fig. 4 also have a higher level direct gap of ≥ 2.7 eV.

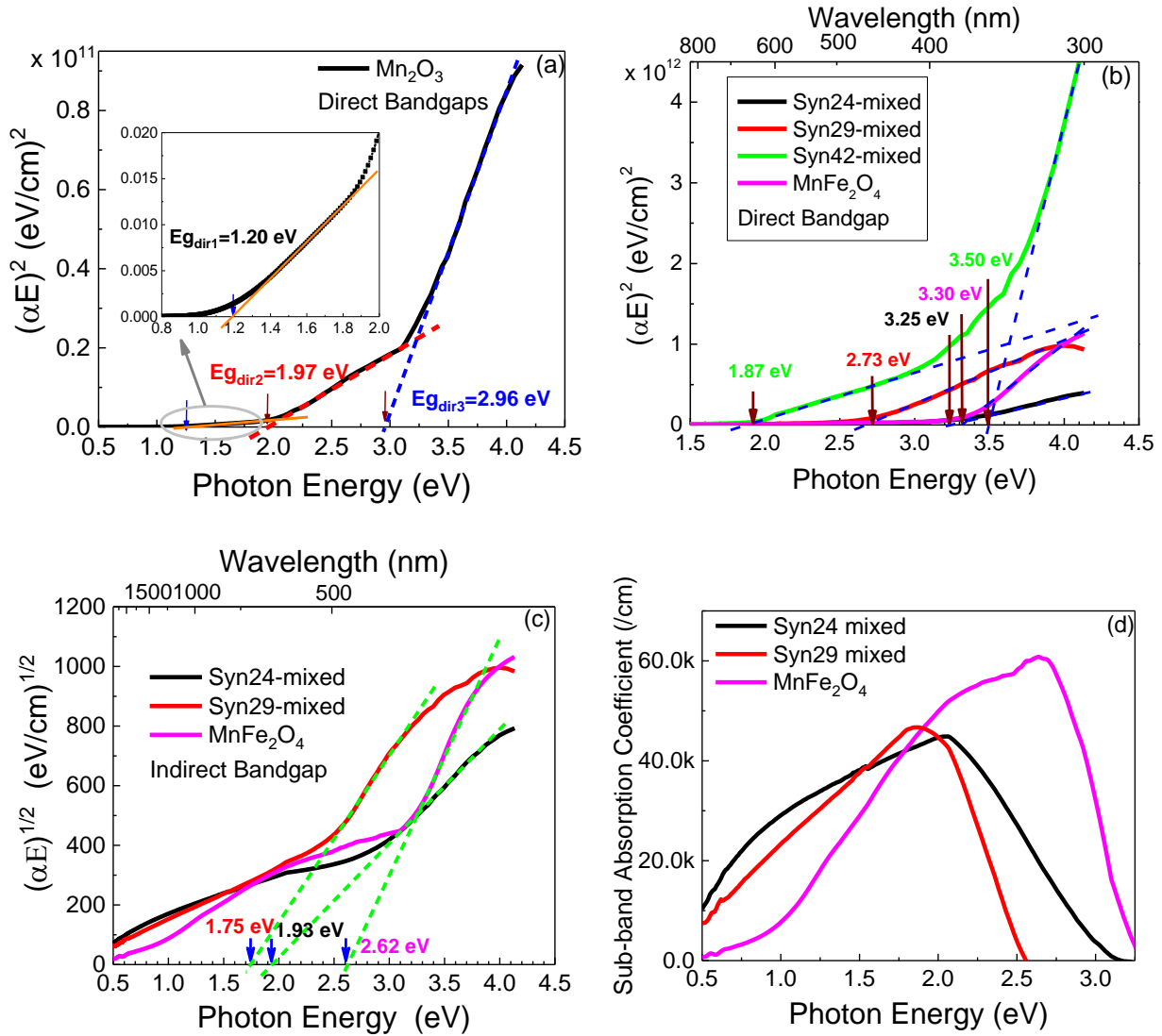

**Fig. 4** Tauc plots of (a) direct bandgaps of $Mn_2O_3$, (b) direct bandgaps of Cu-Mn-Fe oxide Syn24, Syn29, Syn42 and $MnFe_2O_4$ NP materials and (c) indirect band gaps of Syn24, Syn29, $MnFe_2O_4$ NP materials, respectively. (d) Sub-band absorption coefficient spectra of Syn24, Syn29, $MnFe_2O_4$ materials.

Besides band-to-band optical absorption, strong absorption below the indirect bandgap is observed in Syn24 and Syn29 NPs, similar to commercial pure $MnFe_2O_4$ NPs. The sub-bandgap absorption spectra are retrieved by subtracting the fitted band-to-band absorption spectra from the overall absorption spectra for different samples, as shown in Fig. 4(d). The redshift of the sub-band absorption coefficient peak in Fig. 4(d) enables Syn24 and Syn29 coatings to achieve stronger solar absorption in the wavelength range > 800 nm in Fig. 3(a) compared to $MnFe_2O_4$ NP pigmented coating. As will be further discussed later, these sub-bandgap infrared absorption bands are due to d-d transitions of different transition metal cations.

**3.3.2 Retrieving absorption spectra of spinel oxide NPs**: In order to evaluate accurately the optical properties of Cu-Mn-Fe spinel oxide NPs with different compositions, the influence of $Mn_2O_3$ impurity phase is further excluded based on descriptions in Section 3.3 using the absorption spectra in Fig. 4(a) and the weight percentages derived from XRD analyses listed in Table 1. Here we use Syn24, Syn29, and Syn42 as examples to show the trend of transitions from predominantly $CuFe_2O_4$ to mixed $CuFe_2O_4+CuMn_2O_4$ to $CuMn_2O_4$. As shown in Fig. 5(a), $CuMn_2O_4$ spinel oxide NPs in Syn42 has a higher absorption cross section at wavelength <800 nm, while $CuFe_2O_4$ in Syn24 exhibits a relatively strong absorption tail at the wavelength >1200 nm. The absorption cross section curve of mixed $CuFe_2O_4$ and $CuMn_2O_4$ spinels from Syn29 lie between those of $CuFe_2O_4$ and $CuMn_2O_4$. All the synthesized Cu-Mn-Fe spinel oxides including iron-free $CuMn_2O_4$ have larger absorption cross sections (for NP diameter D=50 nm) than $MnFe_2O_4$ NP investigated in our previous work Ref. [19]. Therefore, it is concluded that the pure spinel oxides in the aspect of solar absorption performance are listed in the ascending order as $MnFe_2O_4<CuFe_2O_4<CuMn_2O_4$. On the other hand, the scattering cross sections of pure spinel oxides in Fig. 5(b) differ only slightly, one or two orders of magnitude lower than the corresponding absorption cross sections.

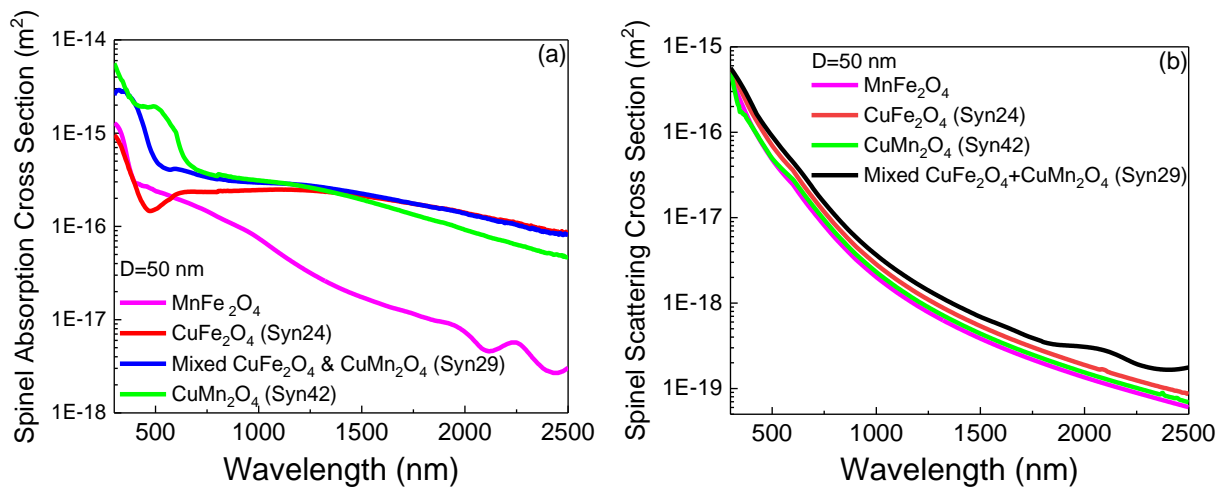

**Fig. 5** (a) Absorption cross sections and (b) scattering cross sections of cubic spinel $MnFe_2O_4$ and Cu-Mn-Fe spinel oxide NPs excluding $Mn_2O_3$ phase from the synthesized Syn24, Syn29 and Syn42. The NP size is 50 nm.

Accordingly, the bandgaps of the spinel phases are refined for synthesized NPs having mixed spinel and $Mn_2O_3$ phases, revealing a slight change from the effective bandgaps incorporating the mixed spinel and $Mn_2O_3$ phases, as summarized in Table 2. For example, the indirect bandgaps of Cu-Mn-Fe oxide spinels in Syn24 and Syn29 are 2.08 eV and 1.74 eV in Fig. 6(a), slightly changing from the corresponding effective indirect bandgaps of 1.93 eV and 1.75 eV in Fig. 4(c), respectively. After excluding the influences of $Mn_2O_3$ taking into account the weight percentage of spinel oxides, the absorption coefficients below the indirect bandgap also become more distinctive, as shown in Fig. 6(b). The sub-bandgap absorption coefficient spectra can be nicely fitted with two Gaussian peaks for the spinel phases in Syn24 and Syn29. Syn24 spinel phase is dominated by $CuFe_2O_4$, and the two peaks at ~1.1 and 2.05 eV correspond to transitions between $Fe^{2+}$ ions and $Fe^{3+}$ ions in octahedral sites through $(Fe^{2+}_{Oh})a_{1g}$-$(Fe^{3+}_{Oh})t_{2g}$ and $(Fe^{2+}_{Oh})a_{1g}$-$(Fe^{3+}_{Oh})e_g$, respectively[34,35]. For Syn29 with mixed $CuFe_2O_4$+$CuMn_2O_4$ spinel phases, these two peaks are shifted slightly to 1.25 and 2.00 eV, most likely due to the d-d transitions of $Mn^{3+}$ on tetrahedral (1.2 eV) and $Cu^{2+}$/$Mn^{3+}$ on octahedral sites, respectively.[28, 36,37] By comparison, the sub-bandgap IR absorption of Syn 42 spinel phase (dominated by $CuMn_2O_4$) only shows a single peak at ~1.1 eV, while the higher energy peak is merged into the direct gap transition at >1.84 eV. This peak is between the tetrahedral site $Cu^{2+}$ absorption band peaked at ~0.8 eV and that of $Mn^{3+}$ peaked at ~1.2 eV, biasing more towards the latter. [28, 36, 37] On the other hand, the peak positions of $MnFe_2O_4$ are blueshifted to 1.75 eV and 2.35 eV for intervalence $Fe^{2+}$-$Fe^{3+}$ transitions. An additional peak at 2.8 eV is ascribed to oxygen and $Fe^{3+}$ ions in octahedral sites for $MnFe_2O_4$. The strong sub-bandgap absorption in Cu-Mn-Fe spinel oxide is beneficial to solar absorption, especially in the near infrared range.

**Table 2.** Bandgaps of synthesized NPs and commercial NPs, and the solar absorptance of NP-pigmented coatings on Inconel substrates.

| NP No. | Spinel Weight Percentage | Effective Eg (eV)** | | Spinel Eg (eV) | | $\alpha_{Solar}$ |
|---|---|---|---|---|---|---|
| | | $Eg_{indir}$ | $Eg_{dir}$ | $Eg_{indir}$ | $Eg_{dir}$ | |
| Syn23 | 100% | 1.94 | 3.17 | 1.94 | 3.17 | 91.0% |
| Syn24 | 46.3% | 1.93 | 3.25 | 2.08 | 3.28 | 96.5% |
| Syn25 | 69.0% | 1.88 | 3.25 | 2.02 | 3.25 | 89.8% |
| Syn26 | 71.2% | 1.65 | 3.16 | 1.66 | 3.13 | 95.4% |
| Syn29 | 30.2% | 1.74 | 2.73 | 1.75 | 2.48 | 95.3% |
| Syn31 | 43.3% | 1.59 | 3.10 | 1.98 | 3.11 | 93.6% |
| Syn33 | 29.8% | 1.61 | 2.94 | 1.83 | 2.80 | 95.7% |
| Syn35 | 82.3% | 1.67 | 2.97 | 1.75 | 2.99 | 93.7% |
| Syn42 | 70.9% | - | 1.87&3.50 | - | 1.84&3.42 | 96.9% |
| $MnFe_2O_4$ | 99.99% | 2.62 | 3.30 | 2.62 | 3.30 | 88.4% |
| $Mn_2O_3$ | 0% | - | 1.20,1.97& 2.96 | - | - | 87.8% |

*EDS measured element ratio of Cu:Fe:Mn

**The calculation of effective bandgaps takes into account $Mn_2O_3$ impurity.

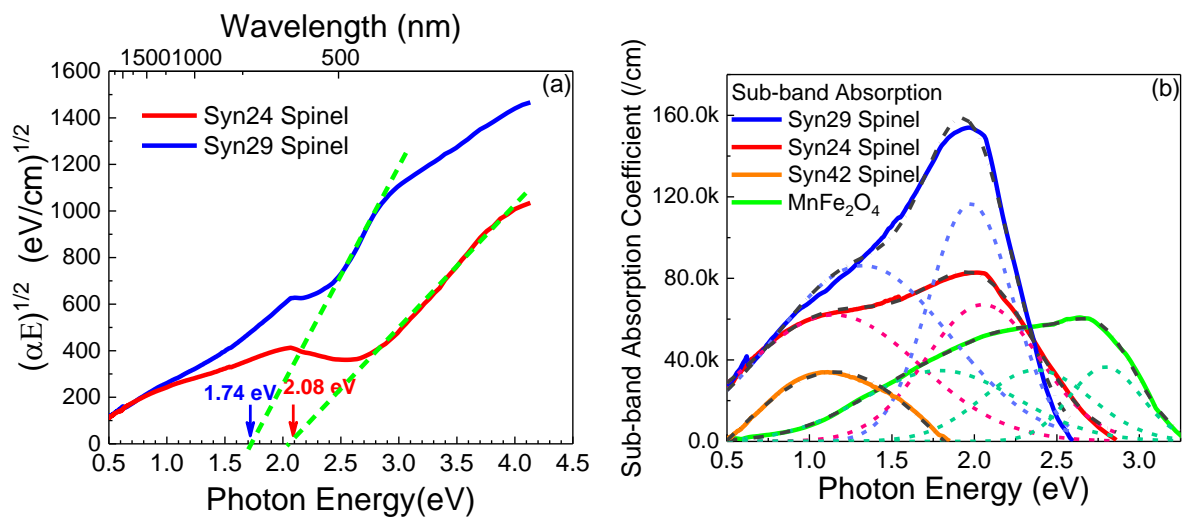

**Fig. 6** (a) Refined indirect bandgaps of Cu-Mn-Fe oxide spinels excluding impurity $Mn_2O_3$ phase from the synthesized Syn24 and Syn29. (b) Deconvolution of sub-band absorption coefficient spectra of spinel phases of Syn24, Syn29, and $MnFe_2O_4$ by Gaussian peak fitting.

**3.3.3 Overall comparison and summary:** After successfully determining and comparing the optical properties of NP spinel materials from their coatings on the quartz substrate for the NPs in Syn24, 29, and 42, we further extend this method to the NPs from all the solar selective coatings. The results of bandgaps and solar absorptance are summarized in Table 2 and plotted in Fig. 7. More details are presented in the Supporting Information, including measured reflectance spectra and coating absorption spectra in Fig. S5, extracted absorption coefficients of spinel oxide NP material in Fig. S6, and wavelength-dependent dielectric function of spinel phases in Fig. S7. Based on the behavior of wavelength-dependent dielectric function, the Cu-Mn-Fe spinel oxide NPs are classified as Mn-substituted $CuFe_2O_4$ group (Syn23, Syn24, Syn25, Syn26, Syn35 ) and Fe-substituted $CuMn_2O_4$ group (Syn29, Syn31, Syn33, Syn42), in agreement with the observations in XRD analysis (Fig. S2(c)).

All the synthesized Cu-Mn-Fe oxide samples have a higher solar absorptance than that of $MnFe_2O_4$ (88.4%) with the same deposition conditions. The best performance are Syn24 and Syn42 NP-coatings with a solar absorptance of 96.5%, 96.9%, respectively. Further considering the data we reported recently in Ref. 29 for Cu-Mn-Cr spinel oxide NPs, we can summarize the following results on modifying the optical properties of $CuMn_2O_4$ spinel NPs by Cr and Fe alloying:

(1) As shown in Fig. 7(a), in the case of Cu-Mn-Fe spinel oxide NPS, the indirect bandgap at ~1.9 eV is almost independent of the Mn/(Mn+Fe) ratio, while the direct bandgap shows a trend of decrease with increasing Mn composition. This correspondingly increases the absorption coefficient in the visible light regime for Mn-rich NPs.

(2) As a result of (2), Fig. 7(b) shows that the solar absorptance generally increases with the Mn/(Mn+Fe) ratio due to the redshift and the correspondingly enhanced direct gap absorption in the visible regime. The relation is almost linear except for a couple of outliers, which may be related to the detailed cationic site occupation of those spinel phases. In contrast, substituting Mn with Cr tends to increase the solar absorptance from ~97% to ~98% [29]. This is an interesting correspondence to the periodic table, where Cr and Fe are neighbors of Mn on either side and modify the solar absorptance in opposite ways.

(3) Similarly, alloying with Cr redshifts the d-d transition IR absorption peak from 1.1 eV to 0.95 eV, while alloying with Fe slightly blueshifts this peak to 1.25 eV, again shifting the IR absorption band to opposite directions. It could be associated with the change in ionic radii upon substitution of Mn ions. These features can be applied to finely tune the IR absorption peak position to further optimize the solar selectivity.

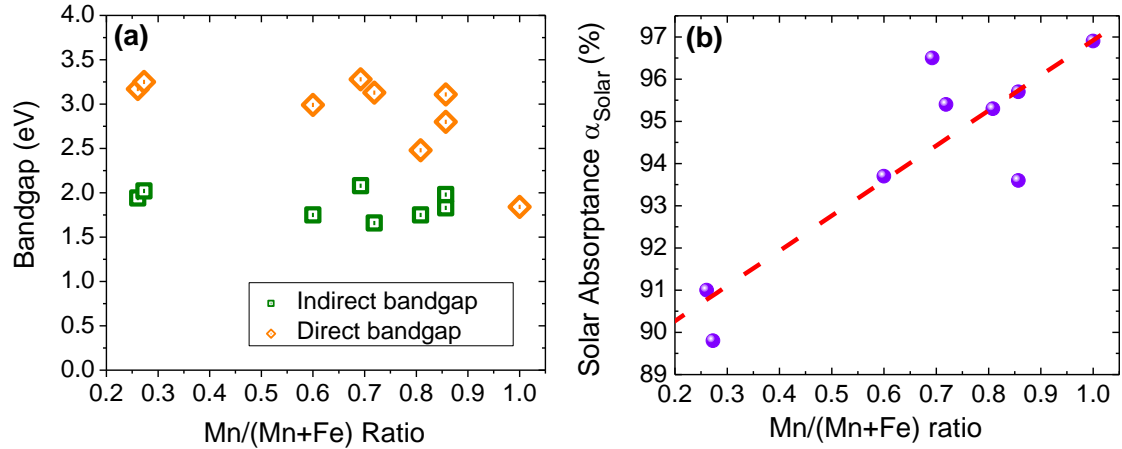

**Fig. 7** (a) Direct and indirect bandgaps vs. Mn/(Mn+Fe) ratio in Cu-Mn-Fe spinel oxide NPs. (b) Solar absorptance of the NP-pigmented solar coatings as a function of Mn/(Mn+Fe) ratio in the NPs.

### 3.4 Computational Optimization of Cu-Mn-Fe Oxide NP-Pigmented Coatings

Based on the optical parameters obtained in Section 3.3, the solar absorptance $\alpha_{solar}$ of synthesized NP-pigmented coatings is optimized using the derived effective absorption and scattering cross sections of NP (D=50 nm) including $Mn_2O_3$ impurity phase. For the calculations of thermal emittance $\varepsilon$ and overall efficiency $\eta_{therm}$, the absorption and scattering cross sections are obtained by extrapolating the refractive index data into the mid IR regime, and the silicone matrix absorption in the IR range is also taking into account using reference silicon coating samples without NPs. As shown previously in Refs. 19 and 29, the latter is the dominant factor in thermal emittance. Since Syn24 and Syn42 offer the highest solar absorptance with the same volume concentration and coating thickness, we only focus on Syn24 and Syn42 here with the corresponding optical data in Fig. 3(d).

Fig. 8 shows color mappings of $\alpha_{solar}$ and $\eta_{therm}$ vs. NP volume concentration $f$ and coating thickness $d$ of Syn24 and Syn42 coatings. Clearly, large regimes with excellent fabrication tolerance are available to offer a high $\alpha_{solar} \geq 97.3\%$ and $\eta_{therm} \geq 94.1\%$ for Syn24 NP-coatings, and $\alpha_{solar} \geq 97.7\%$ and $\eta_{therm} \geq 94.8\%$ for Syn42 NP-coatings, respectively. Theoretically predicted data points corresponding to the experimental parameters of $f$=13 vol. % and $d$=8.5 µm are also indicated with red asterisks in Fig. 8 (a)-(d), in good agreement with the experimentally measured $\alpha_{solar}$ and $\eta_{therm}$ listed in Table 2 and shown later in Fig. 11 ($\alpha_{solar}$ =96.5% , $\eta_{therm}$ = 93.1% for Syn24, and $\alpha_{solar}$ =96.9% , $\eta_{therm}$ = 93.6% for Syn42 ). It further confirms that the derived optical properties of NPs with our approach described in Fig. 1 can successfully optimize the NP-pigmented solar selective coatings. This approach for inverse problem of four-flux-radiative modeling can also deepen the understanding of the optical degradation during thermal cycle testing in section 3.5.

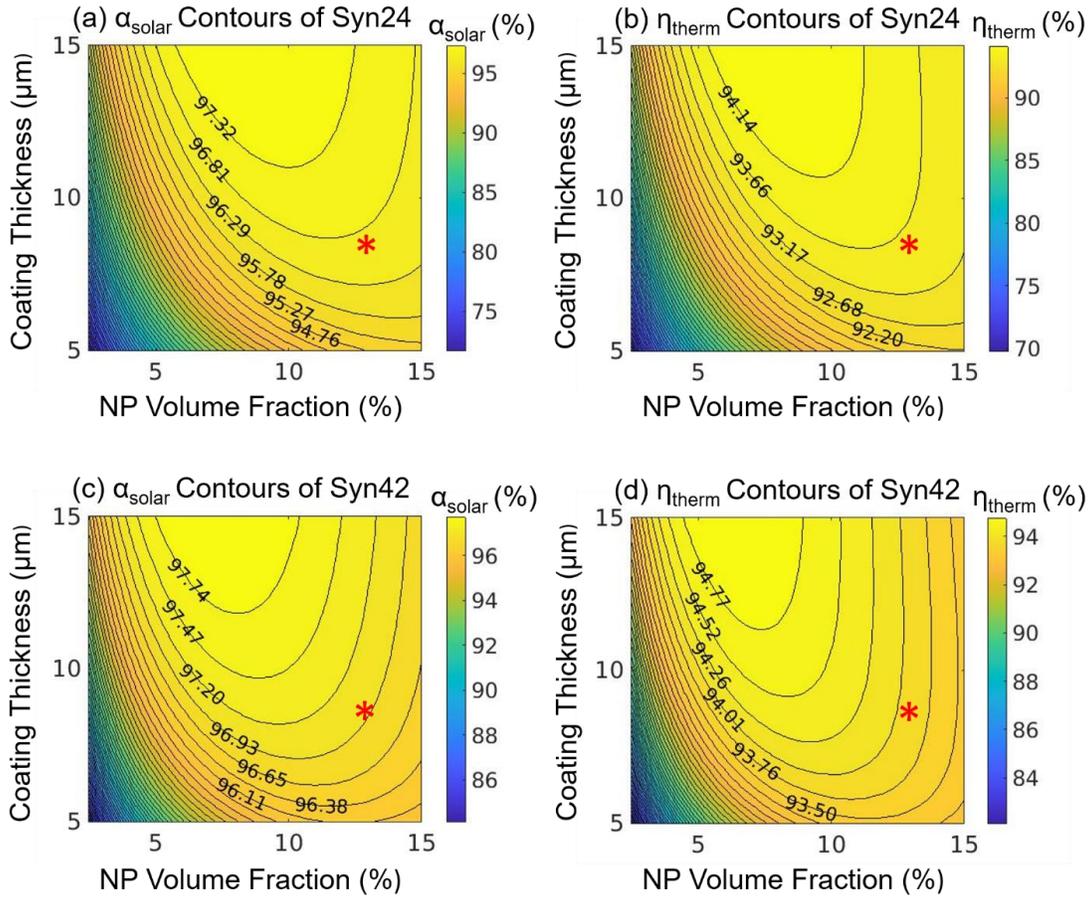

**Fig. 8** Theoretically modelled color mappings of solar absorptance and thermal efficiency vs. NP volume concentration and coating thickness, (a)(b) Syn24 NP-coating, and (c)(d) Syn42 NP-coating. The determined effective scattering and absorption cross sections of Syn24 and Syn42 are used in the four-flux

radiative modeling, respectively. The red stars indicate the NP-pigmented coatings with a NP volume fraction of 13% and a coating thickness of 8.5 µm.

### 3.5 Cu-Mn-Fe Oxide NP-Pigmented Solar Coating Optical Performance and Thermal Cycling Tests

**3.5.1 Thermal stability testing**: We deposited the optimized NP-pigmented coatings (volume fraction $f$=13 vol.%, thickness $d$~8.5 µm) on Inconel for optical measurement and thermal cycling tests at 750°C in air. Each simulated day-night thermal cycle comprises 12 h at 750 °C in air and 12 h cooling to room temperature. The samples are tested for 30-60 day-night cycles. The NPs used in the coatings include Syn23 (dominated by $CuFe_2O_4$), Syn24 (Mn-substituted $CuFe_2O_4$ and 54 wt.% $Mn_2O_3$), Syn26 ($CuFeMnO_4$ and 29 wt.% $Mn_2O_3$), Syn29 (Mn-substituted $CuFe_2O_4$, Fe-substituted $CuMn_2O_4$ and 70 wt.% $Mn_2O_3$ ) and Syn42 ($CuMn_2O_4$ and 29 wt.% $Mn_2O_3$). The thermal stability of commercial $MnFe_2O_4$-coatings can be found in our previous work[19]. SEM and X-ray are used to identify the influences of thermal treatment on micro-morphology and crystalline phases. No cracks, warps or flaking are observed in the coatings on Inconel substrate after long-term thermal stability test. Negligible change is observed in the microstructure of a typical example (Syn24 coating) after 60 simulated day-night cycles between 750°C and room temperature in air, comparing SEM images in Fig. S8 after the thermal cycles with those in Fig. S4 before thermal cycles. XRD data indicate that the Cu-Mn-Fe oxide spinel phases and $Mn_2O_3$ phase are relatively stable. As shown in Fig. S9(a), Syn23 with nearly 100% Fe-rich spinel oxide start to decompose into more stable phases, including $Fe_2O_3$, after 20 thermal cycles, similar to the behavior of $MnFe_2O_4$ after long-term annealing at 750ºC reported in our previous work[19]. By comparison, incorporating $CuMn_2O_4$ greatly enhance the spinel oxide phase stability. For example, Fig. S9 (b) reveals that the change of spinel weight percentage in Syn24 is within 2 wt.% after 60 thermal cycles, and no phase decomposition was observed. This is also the case for $CuMn_2O_4$-dominated spinel oxide NPs in Syn 42. Therefore, increasing Mn contents not only increases solar absorptance as shown in Fig. 7(b), but also enhances the spinel phase stability compared to their Fe-rich counterparts.

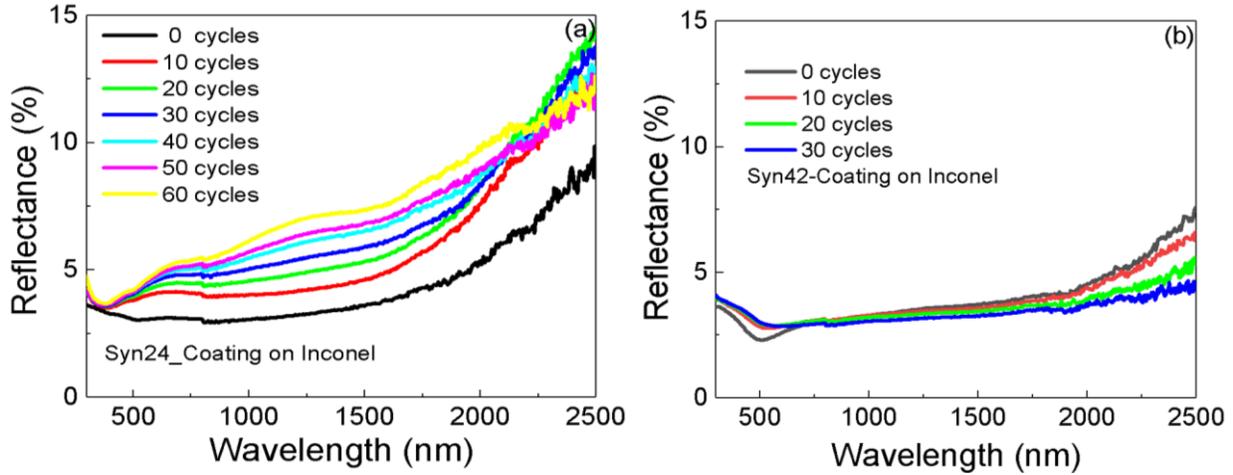

**Fig. 9** Comparison of reflectance spectra in the solar spectral range after different thermal cycles between 750°C and room temperature in air (a) Syn24-coating on Inconel, and (b) Syn42-coating on Inconel. Each simulated day-night thermal cycle comprises 12 h at 750 ºC in air and 12 h cooling to room temperature.

The thermal stability of optical performace can be further revealed in the reflectance spectra of Cu-Mn-Fe oxide NP- coatings after various thermal cycles. The reflectance spectra of NP-pigmented coatings in the solar spectrum range after different thermal cycles are shown in Fig. 9 (Syn24 and Syn42) and Fig.S10 (Syn23, Syn26 and Syn42). The reflectance of Syn24-coating in the solar spectral regime increase largely for the first 10 cycles, and then gradually increase with more thermal cycles. Correspondingly, $\alpha_{solar}$ of Syn24-coating drops by 0.9% from 96.7% to 95.8% after the first 10 cyclic thermal tests, and drops 0.7% after another 20 cycles and 0.5% after 30 cycles. The reflectance spectra of Syn42 are stable and $\alpha_{solar}$ maintains at ~97% vs. the thermal cycles. The optical degradation of Syn24-coating mechanism will be discussed later. On the other hand, the typical reflectance spectra in the infrared range are shown in Fig.S11 (Syn24 and Syn42) for the thermal emittance calculation. The big dip at ~8000 nm in the reflectance curves corresponds to the absorption bands of the silicone resin. It is possible to reduce the thermal emittance loss and improve the solar absorption selectivity by choosing appropriate better silicone materials with less thermal emittance.

Accordingly, the solar absorptance, thermal emittance and thermal efficiency as a function of thermal cycles are summarized in Fig. 10. The trends of thermal efficiency $\eta_{therm}$ are similar to solar absorptance $\alpha_{solar}$, indicating that thermal emittance $\varepsilon_{therm}$ loss is less significant under the sun concentration ratio C=1000 based on Equation 1. In contrast to non-selective Pyromark ($\alpha_{solar}$~97%, $\varepsilon_{therm}$~87%), all synthesized-NP coatings exhibit solar absorption selectivity with the thermal emittance below 60%, which can be further reduced by choosing better low-emittance high-temperature silicone resin. Syn42-coating has the best thermal stability, with negligible degradation in the optical performance after 30 thermal cycles, i.e. $\alpha_{solar}$ ~97%, $\varepsilon_{therm}$ ~55% and $\eta_{therm}$ ~93.5%. These performances satisfy the requirements of the next-generation, high-temperature solar selective coatings ($\alpha_{solar}$ > 95%, and $\varepsilon_{therm}$ < 60%) to ensure LCOE reduction. Coatings consisting of Mn-substituted $CuFe_2O_4$ (Syn24), Fe-substituted $CuMn_2O_4$ (Syn29) or $CuMnFeO_4$ (Syn26) show a similar trend of degradation in the solar absorptance and thermal efficiency with thermal cycles. The thermal efficiency decreases from 93.1% to 90.1% for Syn24 coatings after 60 thermal cycles, from 92.2% to 89.1% for Syn29-coating after 30 cycles, and 91.8% to 90.6% for Syn26-coating after 20 cycles. By contrast, $\alpha_{solar}$ and $\eta_{therm}$ of Syn23-coating with nearly pure Fe-rich $CuFe_{1.7}Mn_{0.6}O_4$ spinel *increases* from 87.8% to 90.2% due to the emergence of crystalline rhombohedral $Fe_2O_3$ (JCPDS no. 33-0664) phase (Fig. S9(a))[38]. This behavior is similar to that of $MnFe_2O_4$ NPs, where thermodynamically driven spinel phase decomposition at 750°C actually increases rather than decrease the thermal efficiency. Overall, the thermal stability of Cu-Mn-Fe oxide NP-coatings at 750°C in the air ranks in the order: Fe-substituted $CuMn_2O_4$ ≈ Mn-substituted $CuFe_2O_4$ ≈ $CuMnFeO_4$ ≪ $CuMn_2O_4$ in terms of optical performances and phase stability. On the other hand, $CuFe_2O_4$ tends to decompose after long-term thermal cycles with ~2% *increase* in thermal efficiency, although still lower than the case of $CuMn_2O_4$. It also suggests that the iron ions should be tightly linked to the performance degradation. This assumption will be confirmed with coating absorption coefficients derived from the reflectance spectra in Fig. 9.

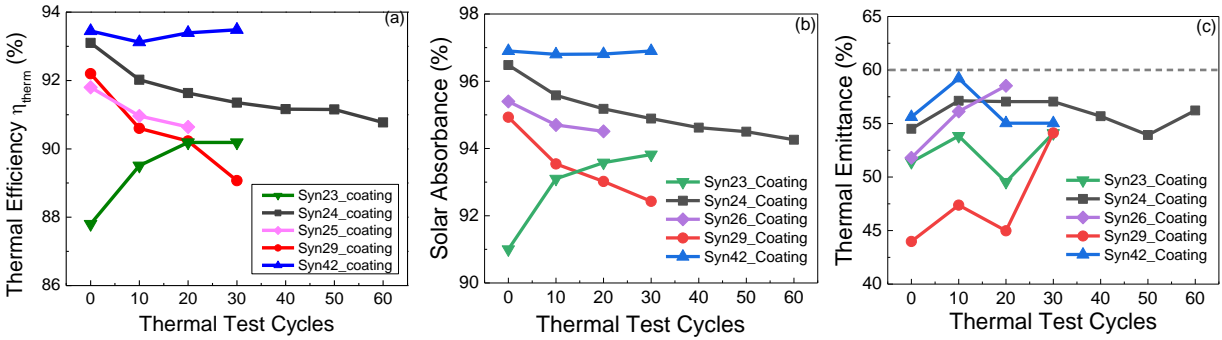

**Fig. 10** Comparison of optical performance of NP-pigmented coatings as function of thermal test cycles at 750°C in the air (a) thermal efficiency, (a) solar absorptance and (c) thermal emittance.

**3.5.2 Degradation mechanism of Cu-Mn-Fe spinel NP-pigmented solar coatings:** In order to understand the degradation mechanism of Syn24-NP coatings, the effective coating absorption coefficients are determined with the same approach described in Section 3.3. Fig. 11 (a) compares the effective NP coating absorption coefficients vs. various thermal treatment cycles. Some interesting features are identified: (1) With increasing thermal cycles, the effective coating absorption coefficient is decreased, where the reduction amount is shown in Fig. 11(b). (2) the reduction is more prominent for the first 10 cycles, i.e., a peak reduction of 0.1/um after 10 cycles, and additional 0.05/um after another 50 cycles; (3) the reduction spans a broad wavelength range between 500 nm and 2000 nm. The similarity between the 10 cycles-solar absorption reduction spectra of Syn24 coating on Inconel and the quartz excludes the substrate influence, such as the oxidation of Inconel substrate. (4) the spectral shape of the reduction in the optical coefficients coincides with that of the d-d transition optical absorption spectrum of Syn24 NP in Fig. 4(d), which is replotted in Fig. 11(b) for comparison. Considering that the sub-band absorption peaks are ascribed to the intervalence transitions between $Fe^{2+}$ ions and $Fe^{3+}$ ions in octahedral sites, the reduction in these peaks indicates the number of Fe ions in octahedral sites decreases.

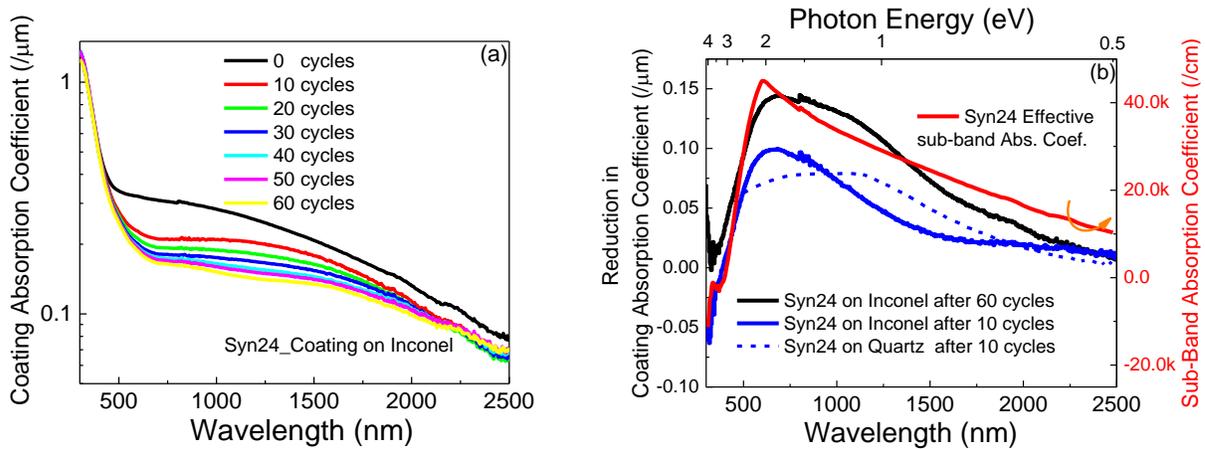

**Fig. 11** (a) Coating absorption coefficients of Syn24-coating on Inconel after various thermal test cycles, (b) Reduction in coating absorption coefficients of Syn24-coating on Inconel after 10 cycles and 60 cycles, and of Syn24-coating on Quartz after 10 cycles. Effective sub-band absorption coefficient of Syn24 is also included for comparison.

X-ray photoelectron spectroscopy (XPS) analysis reveals that $Cu^{1+}$, $Cu^{2+}$, $Mn^{3+}$, $Mn^{4+}$ and $Fe^{3+}$ ions are simultaneously present in $CuFeMnO_4$ in our previous work in Ref. 29. In the case of Cu-Mn-Fe spinel oxides, some $Fe^{2+}$ ions may exist due to partial reduction of $Cu^{2+}$ to $Cu^{1+}$ and oxygen deficiency[39]. Here we examine the thermal-induced cation-redistribution in the tetrahedral and octahedral sites by using the value of the intensity ratio between a pair of diffraction lines of the spinel structure[40,41]. X-ray peak intensity $I_{hkl}$ is proportional to the absolute value square of the structure factors $|F_{hkl}|^2$, multiplicity factor P for the plane *(hkl)*, and the Lorentz polarization factor $L_p$ according to Buerger's equation[40]. Diffraction peaks (220) and (422) are only related to the tetrahedrally-coordinated cations, while (440) peak is attributed not only to cations in tetrahedral sites, but also to those in octahedral sites. Therefore, the intensity ratios of $I_{220}/I_{440}$ and $I_{440}/I_{422}$ are very sensitive to the cationic distribution in the spinel structure[42]. In particular, a higher ratio of octahedral to tetrahedral site occupation for $Fe^{3+}$, $Fe^{3+}(oh)/Fe^{3+}(th)$, will lead to a relative increase in $I_{440}$ compared to $I_{220}$ or $I_{422}$. Although Fig. S9 (b) shows no presence of new phases and negligible change in spinel percentage for Syn24 upon thermal cycling, the intensity ratio $I_{220}/I_{440}$ decreases from ~1.8 to ~1.5, and $I_{440}/I_{422}$ increases from ~1.8 to ~2.2 for Mn-substituted $CuFe_2O_4$ spinel in the Syn24 coating after 60 thermal cycles (Fig. 12). It indicates a rise in the ratio of octahedrally-coordinated to tetrahedrally-coordinated $Fe^{3+}$ ions [38]. Therefore, the most reasonable explanation for the solar absorptance degradation of Cu-Mn-Fe oxide spinels is the oxidation of $Fe^{2+}$ ions in the octahedral sites upon thermal cycles, thus increasing the $Fe^{3+}(oh)/Fe^{3+}(th)$ ratio and suppressing the d-d optical transitions between $Fe^{2+}$ ions and $Fe^{3+}$ ions in the octahedral sites to reduce the optical absorption. This is also consistent with the fact that iron-free $CuMn_2O_4$ and Cu-Mn-Cr oxide NP (Syn42) coatings are both thermally stable upon 750°C/room temperature thermal cycling in air (Ref. 29).

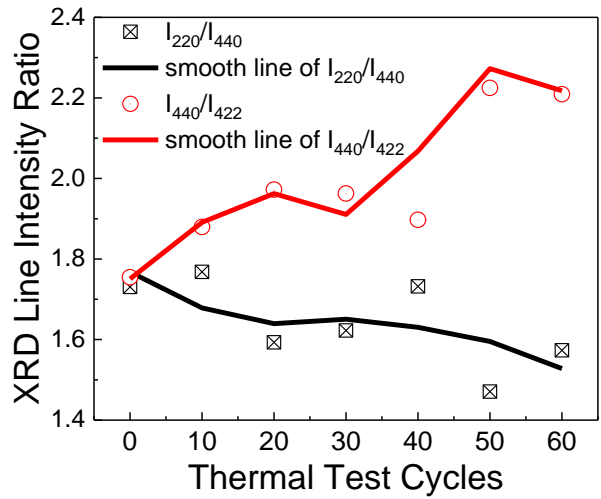

**Fig. 12** XRD Peak intensity ratios of $I_{220}/I_{440}$ and $I_{440}/I_{422}$ as function of thermal test cycles for Mn-substituted $CuFe_2O_4$ spinel in the Syn24-coating.

## 4. Conclusions

Complementing our previous studies on Cu-Mn-Cr spinel NP solar selective coatings [29], in this paper Cu-Mn-Fe spinel oxide NPs are synthesized by co-precipitation method and systematically evaluated for high-temperature solar selective coatings. A reliable, general approach to evaluate the absorption coefficients from the optical measurements of the NP-pigmented coatings is developed based on the inverse problem of the four-flux-radiative method, which agrees very well between theoretical modeling and experiment. The derived absorption properties of NP materials have been directly utilized to elucidate the direct and indirect bandgap transitions in the visible regime as well as the d-d sub-band absorption in the IR regime, thereby optimizing the solar absorptance and evaluating the degradation mechanism of optical performance upon high temperature endurance tests.

Depending on the starting material ratio of Cu:Mn:Fe, the spinel phase may be dominated by $CuFe_2O_4$, $CuMn_2O_4$ or a combination of $CuFe_2O_4$ and $CuMn_2O_4$. The analysis reveals that iron-free $CuMn_2O_4$ spinel is a direct bandgap material with $E_g$=1.84 eV, while the spinel group of Cu-Mn-Fe are fundamentally indirect bandgap between 1.7~2.1 eV. The large sub-band d-d transition absorption peaks for Cu-Mn-Fe oxide spinels are located 1.15~1.30 eV and 2.00~2.05 eV, corresponding to transitions between $Fe^{2+}$ ions and $Fe^{3+}$ ions in octahedral sites. Exactly opposite to Cr alloying, Fe alloying tends to decrease the solar absorptance by blueshifting the direct gap transition in the visible regime as well as the d-d transition peak in the near infrared regime. With the same coating thickness and nanoparticle load, the solar absorptance ranks in the order of $Mn_2O_3$ < $MnFe_2O_4$ < $CuFe_2O_4$ < $CuFeMnO_4$ < $CuMn_2O_4$. On the other hand, the thermal stability at 750°C in air ranks in the order: Fe-substituted $CuMn_2O_4$ < Mn-substituted $CuFe_2O_4$ ≈ $CuMnFeO_4$ ≪ $CuMn_2O_4$.. The significant sub-band absorption is suppressed by the oxidation of $Fe^{2+}$ ions in the octahedral sites upon thermal cyclic tests, causing the degradation in the solar absorptance of Cu-Mn-Fe oxide NP.

The systematic analysis and computational modeling enables us to optimize iron-free $CuMn_2O_4$ sample to demonstrate a high solar absorptance of 96.9%, low emittance of 55.0%, and a high thermal efficiency of ~93.5% under 1000x solar concentration at 750ºC in air, in good agreement with the optimized values calculated with derived optical parameters. Furthermore, $CuMn_2O_4$-NP coating maintains its 93.5% thermal efficiency after long-term thermal cycling between 750°C and room temperature. This solar coating deign and optimization approach can be readily extended to other material systems to speed up the searching and optimize high-temperature pigmented-solar selective coatings.

**Author Contributions**

The authors make equal contributions to the paper. Xiaoxin Wang conducted the optical modeling and data analysis, and drafted the manuscript. Can Xu fabricated the NP-pigmented coatings and performed the thermal endurance testing as well as materials and optical characterization. Jifeng Liu proposed the original idea of investigating Cu-Mn-Fe oxide NP pigments and supervised the project. All authors contributed to the final editing of the paper.

**Acknowledgement**

This project was funded by U.S. Department of Energy, Solar Energy Technologies Office, under the award numbers DE-EE0007112 and DE-EE0008530.

**Supporting information**

Supplementary data associated with this article can be found in the online version at :

**Data Availability**

The raw data required to reproduce these findings are available upon request to the correspondence author at Jifeng.Liu@dartmouth.edu. The processed data required to reproduce these findings are also available upon such requests.

# Supporting Material

# Extracting optical properties of Cu-Mn-Fe spinel oxide nanoparticles for optimizing air-stable, high-efficiency solar selective coatings

Xiaoxin Wang*[†], Can Xu, [†] and Jifeng Liu**

**1 Characterization of nanoparticles (NPs) and NP-pigmented coatings**

**1.1 TEM Data**

Fig.S1 shows ypical TEM images of some as-synthesized NPs with additional information provided by the selected-area electron diffraction（SAED). The average NP sizes are 41.5 nm, 49.5 nm, 49.2 nm, 53.8 nm, and 48.0 nm for as-synthesized Syn23, Syn24, Syn25, Syn26 and Syn29 in Fig.S1(a)-(d), respectively, with a standard deviation of ~10 nm. As-synthesized Syn25 and Syn23 share similar SAED patterns presented in Fig.S1(e), in which the rings represent different diffraction lines of cubic spinel nanocrystals. As-synthesized Syn26 has dominant diffraction rings of cubic α-$Mn_2O_3$ crystals in Fig.S1(f), similar to the case of Syn29.

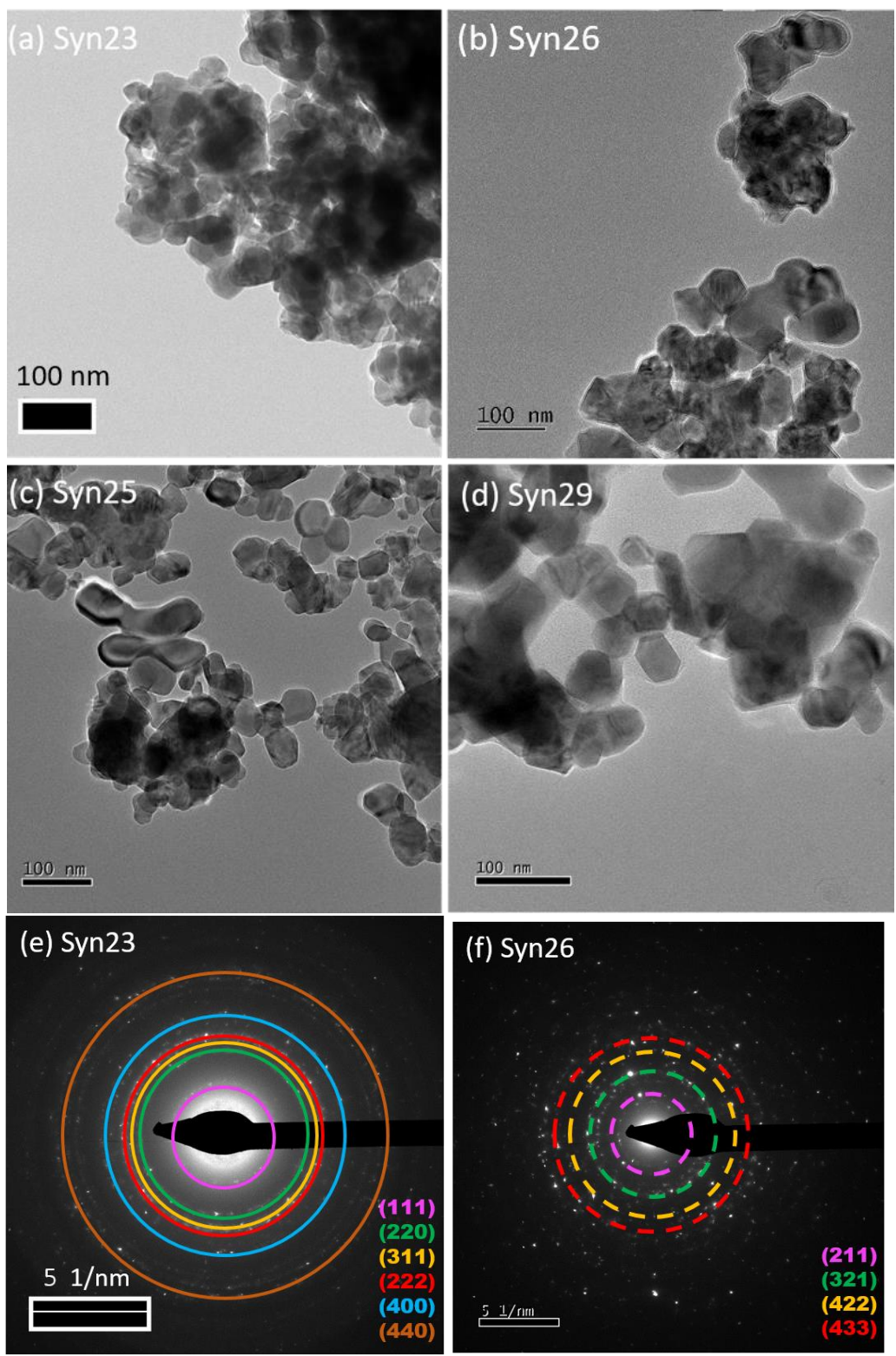

**Fig.S1** TEM images of as-synthesized NPs, (a) Syn23, (b) Syn26, (c) Syn25, and (d) Syn29. Selected area electron diffraction (SAED) patterns of (e) as-synthesized Syn23, cubic spinel Cu-Mn-Fe oxide phase, and (f) as-synthesized Syn26, $Mn_2O_3$ phase.

## 1.2 Supplemental XRD Data

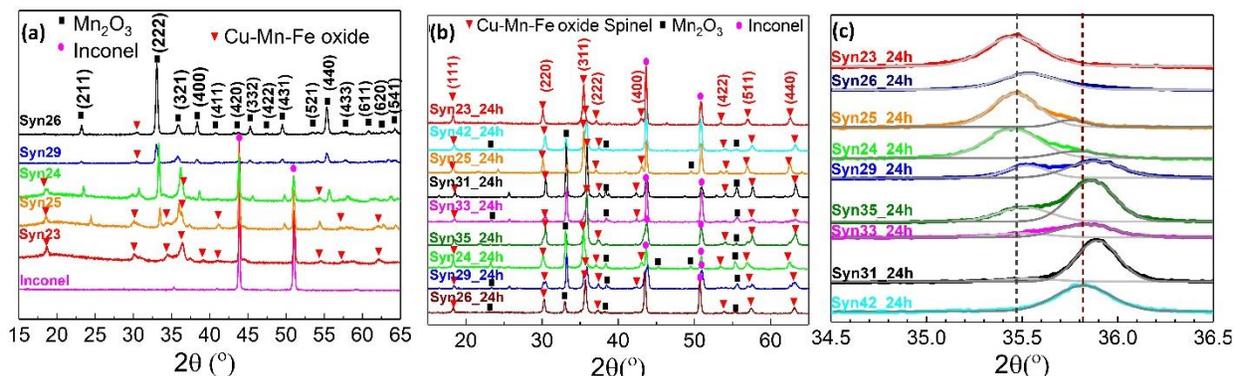

**Fig. S2** (a) XRD spectra of as-synthesized NPs (b) XRD spectra of NPs annealed at 750°C for 24 hours in air. (c) further deconvolution of (311) peaks of annealed samples listed in (b). The peaks at ~35.5° and ~35.8° correspond to $CuFe_2O_4$ and $CuMn_2O_4$, respectively.

The crystalline phases are further identified with the X-ray diffraction technique. Fig. S2(a) compares XRD data of some as-synthesized nanoparticles. As discussed in the main text, a secondary $Mn_2O_3$ phase is inevitably formed along with cubic spinels when the ratio Cu/Mn ≤1.2 due to thermodynamic phase equilibrium. As expected, as-synthesized Syn23 (Cu:Mn:Fe=1:0.5:1.5) is a pure spinel, Syn24 (Cu:Mn:Fe=1:3:1) has a small amount of $Mn_2O_3$, while both Syn29 (Cu:Mn:Fe=1:4:1) and Syn26 (Cu:Mn:Fe=1:6:2) have a dominant $Mn_2O_3$ phase. Higher temperature is required to thermodynamically promote the formation of a stable cubic spinel phase from the co-precipitated nanocrystals, particularly in the cases of low ratio of Cu/Mn < 0.3. Therefore, all the as-synthesized nanoparticle samples or as-deposited coatings are post-annealed at 750°C for 24 h to enhance the spinel oxide phases, as shown in Fig. S2(b). The weight percentages of crystalline phases are quantified by applying a standard intensity correction based on the reference intensity ratio (RIR) from the sample to corundum ($I/I_c$, 4.50 for cubic $Mn_2O_3$ and 5.13 for cubic $CuFe_2O_4$ [S1]), as included in Table 1 of the main text.

Furthermore, when XRD (311) peaks of cubic Cu-Mn-Fe oxides are examined closely (Fig. S2(c)), it consists of two peaks for most annealed samples, which are ascribed to $CuFe_2O_4$ (~35.5°) and CuMn$_2$O$_4$ (~35.8°), respectively. The Fe-substituted CuMn$_2$O$_4$ or Mn-substituted CuFe$_2$O$_4$ leads to a slight shift in the diffraction peak positions. The lattice parameters of Cu-Mn-Fe oxide spinels determined from the diffraction peaks in Fig. S2(b) are listed in Table 1. The value of the lattice parameter for cubic spinel CuFe$_2$O$_4$ system varies from 8.37~8.39Å, consistent the reported value of 8.384 Å[s2]. The lattice parameter of CuMn$_2$O$_4$ is in the range of 8.29~8.31Å, less than that of CuFe$_2$O$_4$. The degree of tetrahedral vs. octahedral inversion or stoichiometry of the cations may also affect the spinel lattice parameter.[s3]

The crystal sizes of the mixed oxide phases are also derived via the Scherrer formula from the deconvoluted (311) spinel peaks and Mn$_2$O$_3$ (222) peak in Fig. S2(c), respectively. As shown in Table 1 of the maintex, Mn$_2$O$_3$ NPs exhibit a large particle size ~50 nm, while spinel Cu-Mn-Fe oxides have a smaller size 30~40 nm. The difference in NP sizes of spinel and α-Mn$_2$O$_3$ phases can explain the relatively large size deviation (~10 nm) observed in the mixed NPs (Fig. S1).

### 1.3 Raman Data

Fig. S3 shows the Raman spectra of the Syn24 and Syn42 on quartz taken in the range from 150 cm$^{-1}$ to 750 cm$^{-1}$. Five characteristic bands (A$_{1g}$+E$_g$+3 F$_{2g}$) are assigned to the cubic inverse-spinel copper ferrite[3]. Specifically, 200 cm$^{-1}$, 466 cm$^{-1}$, and 558 cm$^{-1}$ are originated from three F$_{2g}$ modes, while 258 cm$^{-1}$ and 655 cm$^{-1}$ correspond to E$_g$ and A$_{1g}$ modes, respectively. The appearance of an extra spinel peak at 390 cm$^{-1}$ is associated with a breakdown of the momentum conservation rule due to nanoscale size. Another set of Raman active optical modes are related to cubic Mn$_2$O$_3$( bixbyite) phase, which are located at 314 cm$^{-1}$, 490 cm$^{-1}$, 613 cm$^{-1}$.[4] For Syn42 NPs, the signals of Mn$_2$O$_3$ are weak and the vibration modes of CuMn2O4 have a slight redshift compared to CuFe2O4. So the Raman data provides further evidence that annealed Syn24 has mixed phases of CuFe$_2$O$_4$ and Mn$_2$O$_3$, while Syn42 has a dominant CuMn2O4 phase.

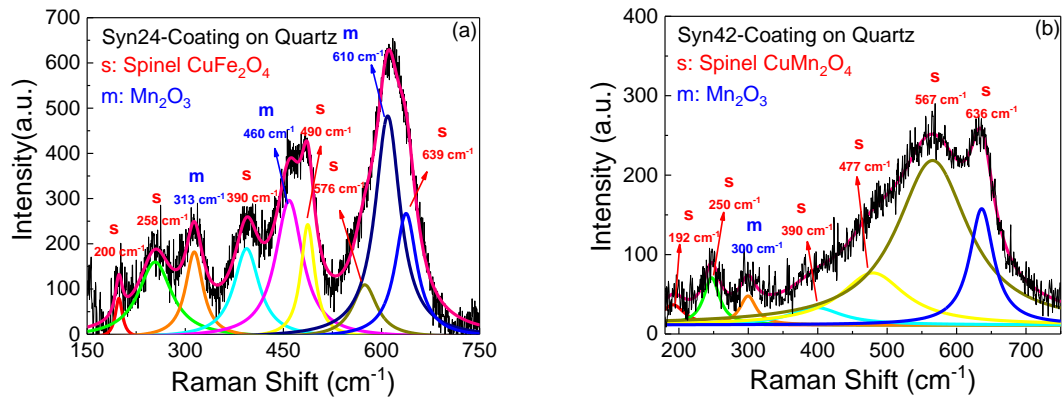

**Fig.S3** Raman spectra of Syn24 and Syn42 NP-pigmented coating deposited on quartz and annealed at 750°C for 24 hours in air.

**1.4 SEM Images of the Coatings before Thermal Cycling**

Fig. S4 (a)-(b) show the SEM images of Syn24-pigmented coating on Inconel. No cracks or warps are found on the coating surfaces. Coating thickness is ~ 8.5 µm estimated from the tilted view of cross section image in Fig. S4(c). From the starting weight ratio of NPs and solid content of the silicone, the volume concentration of NP is calculated as 13%. These two parameters will be used to determine the optical properties of Cu-Mn-Fe oxides.

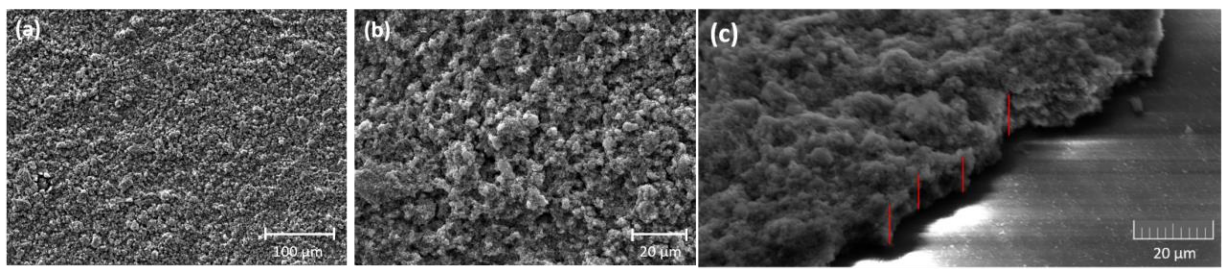

**Fig.S4** SEM images of Syn24 NP-pigmented coatings on Inconel substrate after annealing at 750 °C for 24 hours, (a)-(b) surface microstructure, and (c) cross-section.

## 2. Extracted Optical Properties of NP-pigmented Solar Coatings on Inconel

After successfully extracting the optical properties of some NP spinel materials from their coatings on the quartz substrate, we further extend this self-consistent method to the realistic solar selective coatings on high-temperature alloy substrates such as Inconel. It is also found that the difference in the optical properties (such as refractive index) extracted from the NP coatings (Syn24, Syn29 and Syn42) on quartz and their coatings on Inconel are negligible.

Fig. S5(a) summarize the reflectance spectra of all synthesized NP- solar selective coatings on the high-temperature Inconel substrate. With the same spray coating conditions, the coating thickness is about 8.5 μm and NP volume concentration is 13%. One prominent feature of pigmented coatings is that the reflectance or solar absorptance can be tuned by the choice of NP materials. Based on the solar absorption performance, these coatings can be classified into three groups. Syn24, Syn33 and Syn42 NPs enable lowest reflectance <10% in the full solar spectrum, while the reflectance of Syn23 and Syn25 NP-pigmented coatings continue increasing with the wavelength and reach to >25% at λ=2500 nm. The curves of Syn26, Syn29, Syn31 and Syn35 NP-coatings lie between two groups above-mentioned, but they exhibit different trends. The reflectance of Syn35 and Syn31 rolls up slowly from 5% to 10% and 15%, respectively. Syn29 and Syn26 NP-coatings have a low reflectance ~5% for λ<1200 nm and a gradual increase in reflectance up to 15% at λ=2500 nm. Overall, all NP-pigmented coatings have 95% absorption for λ<500 nm but have a pronounced difference in the solar absorption for λ>500 nm. The distinguished behavior in optical absorption gives rise to various solar absorptance values, as shown in Table 2. All the synthesized Cu-Mn-Fe oxide samples have a higher solar absorptance than that of $MnFe_2O_4$ (88.4%) with the same deposition conditions. The best performance are Syn24 and Syn42 NP-coatings with a solar absorptance of 96.5%, 96.9%, respectively.

The derived effective absorption coefficients of coatings on Inconel are shown in Fig. S5(b). Syn23 and Syn 25 NPs have a lower effective coating absorption coefficient, while Syn42 and Syn29 show the highest value, again indicating that $CuMn_2O_4$ is preferred over $CuFe_2O_4$ in terms of solar absorption. Following the same procedure described in Section 3.3, the information of direct and indirect bandgaps of effective synthesized NP with different compositions is listed in Table 2. Besides a large direct bandgap ranging from 2.9 eV to 3.3 eV, the synthesized NPs (except iron-free Syn42 NP) have a fundamental effective indirect band gap (1.59~1.94 eV).

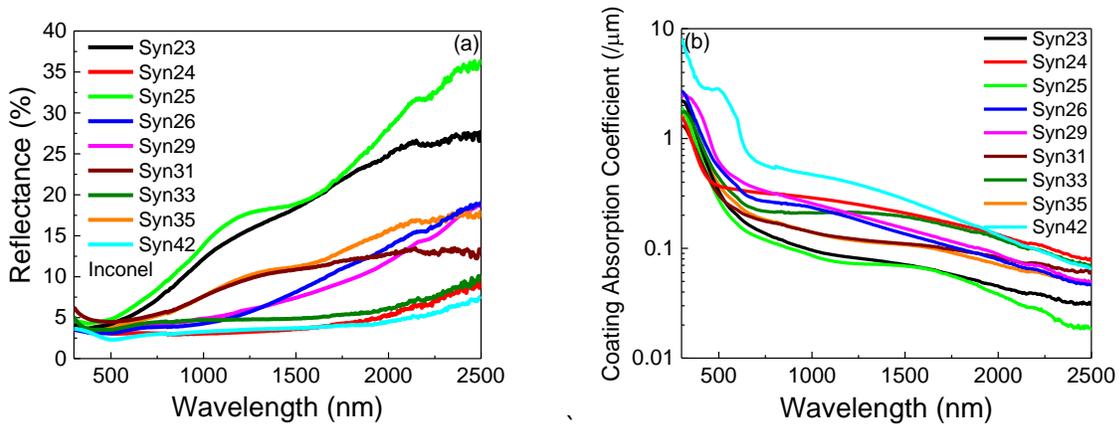

**Fig. S5** Comparison of (a) measured reflectance spectra and (b) derived effective absorption coefficients of synthesized NP-pigmented coatings on Inconel substrates.

Considering a large variation in the spinel weight percentage among those synthesized NPs listed in Table 2 of the maintext, further efforts are made to extract the optical properties of pure spinel phases excluding $Mn_2O_3$ phase influence. Fig. S6 compares the absorption coefficients of spinel NP materials, which are also used to obtain the bandgap information (Table 2). The indirect band gaps of Cu-Mn-Fe oxide spinel phase shows a slight increase, in the range of 1.67~2.08 eV. Fig.S6 also suggests that with the same coating thickness and nanoparticle load, the solar absorptance of the pure spinel oxides ranks in the order of $MnFe_2O_4$ (purchased) < $CuFe_2O_4$ (in Syn23) < CuFeMnO4 (in Syn26) <CuMn2O4 (in Syn42). Specifically, CuMn2O4 material (in Syn42) has a large absorption coefficient at wavelength < 1000 nm (photon energy >1.25 eV), while Fe-substituted CuMn2O4 from Syn33 and Mn-substituted CuFe2O4 from Syn24 have a broad low-energy absorption peak (photon energy <1.25 eV). The mixed CuFe2O4 and CuMn2O4 spinels from Syn29 exhibit a higher absorption coefficient almost in the whole solar spectrum than CuMn2O4 spinel from Syn42. However, the spinel percentage is only 30% in Syn29 vs. 70% in Syn42, so 95.3% for Syn29 NP-coating vs. 96.9% for Syn42 coating, as listed in Table 1. The optical transitions contributing to the absorption coefficients of Cu-Mn-Fe oxide spinels are well presented with the dielectric permittivity $\varepsilon$ ($=\varepsilon_1+i\varepsilon_2$) as function of photon energy in Fig.S8.

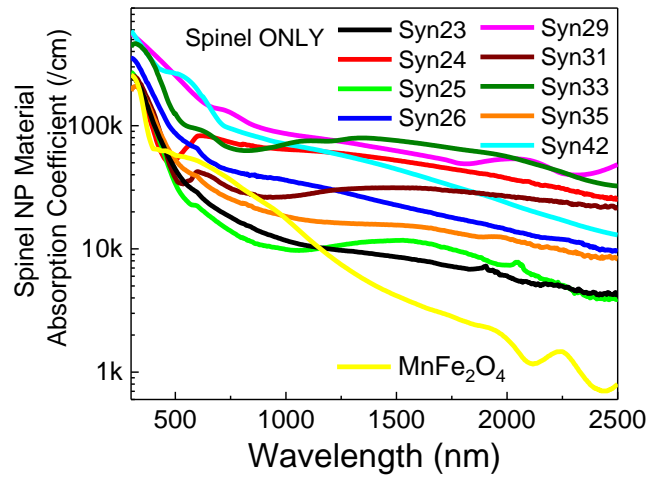

**Fig. S6** Absorption coefficients of spinel materials in synthesized NPs and commercial MnFe2O4.

Based on XRD analysis (Fig.S2(c)) the Cu-Mn-Fe oxide spinels are classified as CuFe2O4 group (Syn23, Syn24, Syn25, Syn26 ) and CuMn2O4 group (Syn31, Syn33, Syn42). This is largely consistent with the trend of wavelength-dependent dielectric function shown in Fig. S7. Among NPs with mixed CuFe2O4 and CuMn2O4 spinels, the $\varepsilon$ trend of Syn35 spinels is more like CuFe2O4, while Syn29 spinels are more like dielectric permittivity behavior of CuMn2O4. Two major features in Fig. S7 is the low-energy absorption for photon energy < 2 eV and high-energy absorption > 2.5 eV. Substitution with Mn in $CuFe_2O_4$ or substitution with Fe in $CuMn_2O_4$ alters the composition and inversion status of oxide spinels, thus changing the optical transition intensity. For example, the spinels of Syn29 and Syn33 have relatively strong low-energy transitions. As discussed in Section 3.2, the low-energy transitions ≤2 eV are assigned to the intervalance transtions between ions in octahedral sites, particularly $Fe^{2+}$ and $Fe^{3+}$ ions. The high-energy absorption > 2.5 eV is ascribed to the transitions between oxygen and ions in octahedral sites.

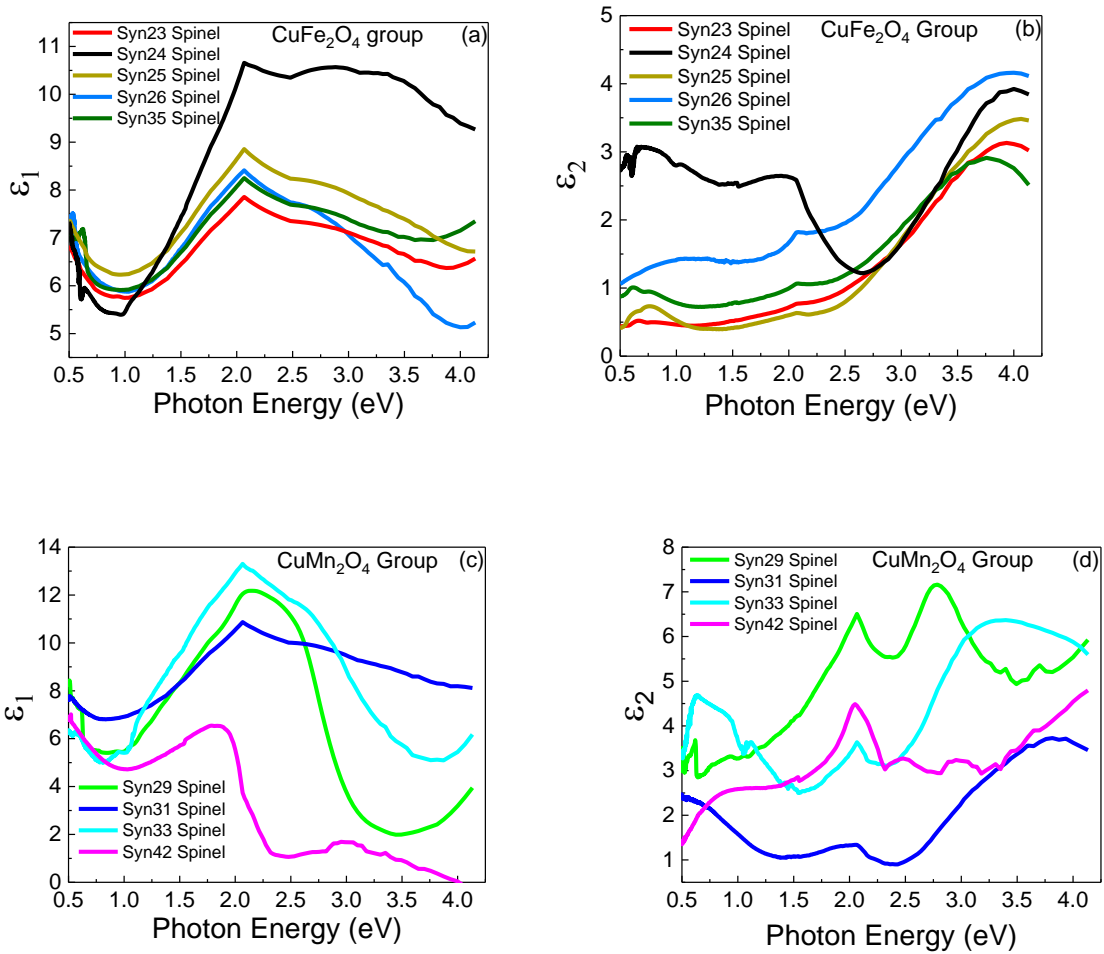

**Fig. S7** Dielectric constants of spinel phases of synthesized NPs as function of photon energy, (a)(b) $CuFe_2O_4$ group, (c)(d) $CuMn_2O_4$ group. Real part $\varepsilon_1 = n^2 - k^2$, and imaginary part $\varepsilon_2 = 2nk$.

**3 High-temperature thermal stability of Cu-Fe-Mn oxide NP-pigmented coatings**

Fig. S8 reveals the surface microstructure of the coatings after 60 thermal cycles testing. No cracks , warps or flakes are observed in the coatings.

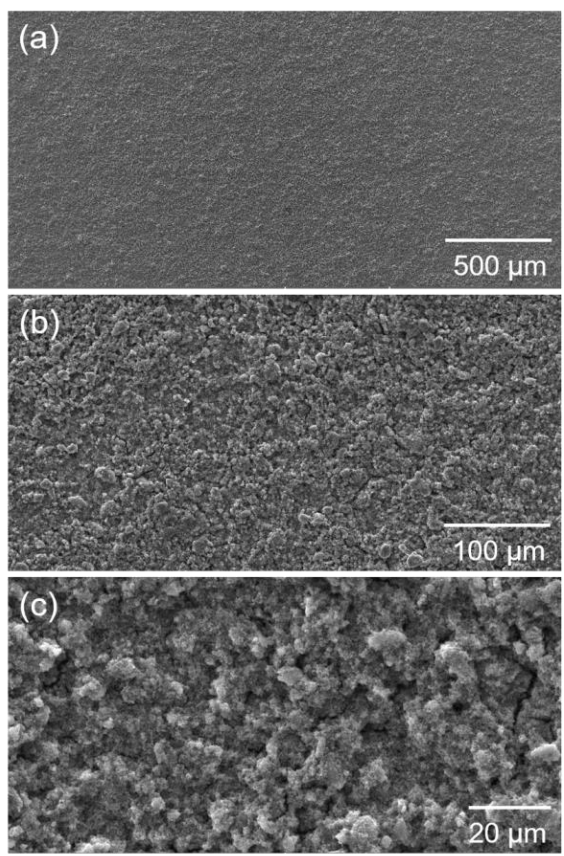

**Fig. S8** SEM images of Syn24 NP-pigmented coatings on flat Inconel substrate after 60 thermal cycles test (720h) at 750°C in the air.

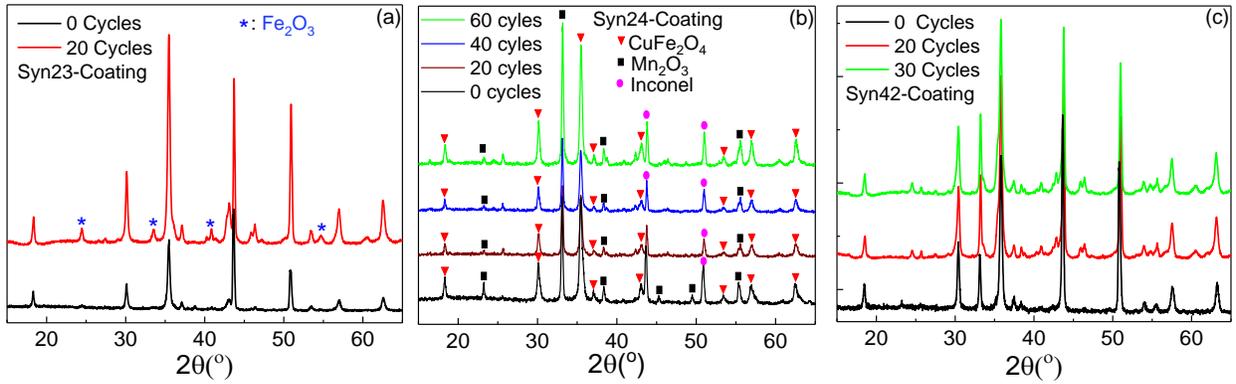

**Fig. S9** XRD spectra of (a) Syn23-coating, (b) Syn24-coating, and (c) Syn42-coating on Inconel substrates after various numbers of thermal cycles between 750 °C and room temperature in air.

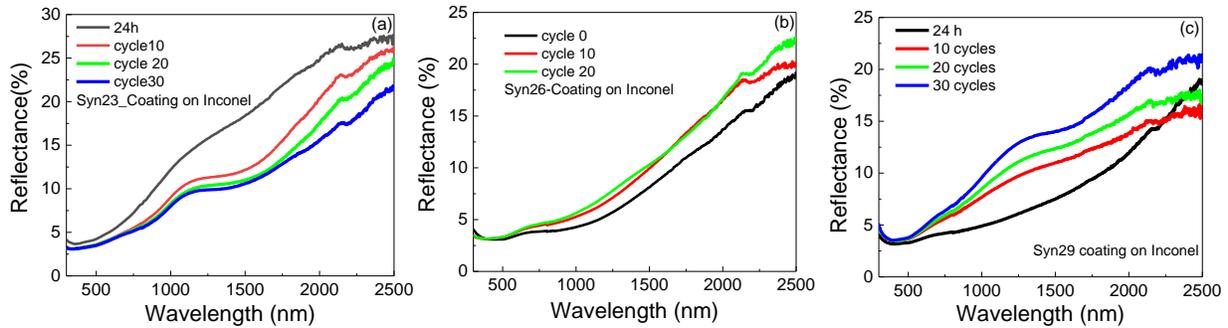

**Fig. S10** Comparison of reflectance spectra of the NP-coatings on Inconel after various thermal test cycles, (a) Syn23 NPs-coating, (b) Syn26 NPs-coating, and (c) Syn29 NPs-coating.

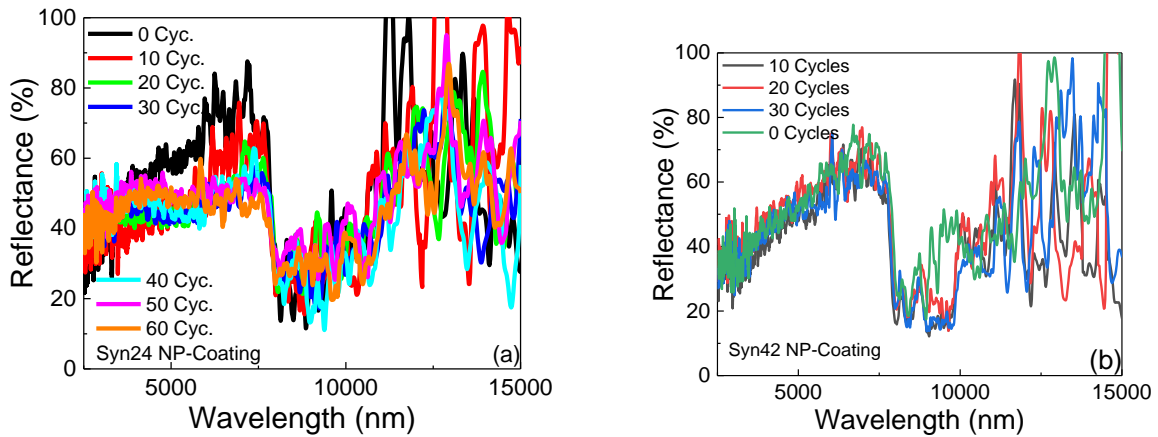

**Fig. S11** Comparison of reflectance spectra in the infrared range for thermal emittance calculation. (a) Syn24 NP coating, and (b) Syn42 NP coating.